\documentclass[%
 reprint,
showkeys,
 amsmath,amssymb,
 aip,pop,
]{revtex4-1}

\usepackage{graphicx}
\usepackage{dcolumn}
\usepackage{bm}
\usepackage{enumerate}
\usepackage{sidecap}

\usepackage{color}

\def\apj{ApJ.}

\def\aap{A\&A}

\def\jgr{J. Geophys. Res.}
\def\grl{Geophys. Res. Lett.}

\def\apjl{Astrophys. J. Lett.}
\def\araa{Annual Review of Astronomy \& Astrophys.}

\begin{document}

\preprint{APS/123-QED}

\title{Quantifying three dimensional reconnection in fragmented current layers} 

\author{P. F. Wyper}
 \email{peter.f.wyper@nasa.gov}
\affiliation{%
Goddard Space Flight Center, 8800 Greenbelt Rd, Greenbelt, MD 20771} 
\author{M. Hesse}%
 \email{michael.hesse-1@nasa.gov}
 \affiliation{%
Goddard Space Flight Center, 8800 Greenbelt Rd, Greenbelt, MD 20771}


\date{\today}

\begin{abstract} 
There is growing evidence that when magnetic reconnection occurs in high Lundquist number plasmas such as in the Solar Corona or the Earth's Magnetosphere it does so within a fragmented, rather than a smooth current layer. Within the extent of these fragmented current regions the associated magnetic flux transfer and energy release occurs simultaneously in many different places. This investigation focusses on how best to quantify the rate at which reconnection occurs in such layers. An analytical theory is developed which describes the manner in which new connections form within fragmented current layers in the absence of magnetic nulls. It is shown that the collective rate at which new connections form can be characterized by two measures; a total rate which measures the true rate at which new connections are formed and a net rate which measures the net change of connection associated with the largest value of the integral of $E_{\|}$ through all of the non-ideal regions. Two simple analytical models are presented which demonstrate how each should be applied and what they quantify. 
\end{abstract}

\maketitle


\section{Introduction}
The process of magnetic reconnection underpins our understanding of many astrophysical phenomena. Examples include solar flares, geomagnetic storms and saw tooth crashes in tokamaks \cite{Zweibel2009,PriestForbes2000}. Yet a complete understanding of this enigmatic plasma process remains illusive, despite decades of research.

Fundamentally, magnetic reconnection is the process whereby excess energy in a magnetic field is liberated by the reorganization of a magnetic field's connectivity in the from of plasma heating, bulk fluid motions and particle acceleration. Classically, this is envisioned to occur in a single well defined region of high electric current, within which non-ideal effects dominate and the plasma becomes decoupled from the magnetic field \cite{Parker1957,Sweet1958,Petschek1964}. However, in recent years the importance of instabilities which fragment reconnection regions has been more fully appreciated. In particular, in two dimensions high aspect ratio current sheets have been shown to be highly unstable to tearing with the resulting dynamics dominated by the formation and ejection of magnetic islands \cite{Loureiro2007,Bhattacharjee2009}, whilst 3D simulations have emphasized the importance of flux rope formation, braiding and the possible development of turbulence \cite{Daughton2011,Wyper2014a,Pontin2011b}. Observations of plasma blobs and bursty radio emissions in the extended magnetic field beneath erupting CME's as well as bursty signatures of reconnection in the Earth's magnetotail appear to somewhat corroborate this picture \cite{Karlicky2014,Liu2010,Hesse1998,Shay2003}.

An important diagnostic of any reconnection scenario is the rate at which the process occurs. In two dimensions reconnection occurs only at X-points, with the rate of reconnection given simply by the electric field at this position. If the current layer is fragmented then the only topologically stable situation is one in which only a single X-point resides at the boundary between the global flux domains. The reconnection rate is then the electric field measured at this ``dominant" X-point (e.g. \citet{Wyper2014b}).

In three dimensions (3D) the picture is more complex. When reconnection involves 3D nulls, separatrix surfaces divide up the magnetic field into regions of differing connectivity. The rate of reconnection can then be defined as the flux transfer across these surfaces \cite{Pontin2005}, or past separators which sit at the intersection of different separatrix surfaces \cite{Vasyliunas1975}. If the non-ideal regions spanning the separatrix surfaces are fragmented then considering flux transfer across segments of a separatrix surface \cite{Wyper2013b,Wyper2013} or along multiple separators \cite{Parnell2008} if they exist allows the reconnection rate to be quantified. Unlike 2D, where X-points other than the dominant X-point do not directly contribute to the reconnection rate (although they may indirectly affect it), in 3D reconnection across a separatrix surface in multiple places or at multiple separators all contribute towards the total rate of flux transfer between the main topological domains. This leads to the surprising result that in 3D {\it{two}} measures of reconnection may be used when reconnection occurs in fragmented current layers. One that measures the total rate at which flux is reconnected (taking account of recursively reconnected magnetic flux) and a net measure of the combined effects of each of the fragmented non-ideal regions. The former is the true reconnection rate for any problem, but the latter may be of interest when the large scale effects of a reconnection site are being considered.

Furthermore, in 3D reconnection may also occur in the absence of magnetic null points. In this case the lack separatrix surfaces against which reconnection can be defined requires a more general approach to the problem. The theory of General Magnetic Reconnection (GMR) encompasses reconnection across separatrices \cite{Schindler1988,Wyper2013b} as well as describing reconnection in situations without them. The theory of GMR has shown that for a single isolated non-ideal region the rate of reconnection is given by the maximum of $\int{E_{\|}dl}$ on all field lines threading the non-ideal region \cite{Hesse1988,Hesse1993,Schindler1988}. However, the question remains as to how to measure reconnection in fragmented current layers without the presence of separatrix surfaces or in situations where separatrix surfaces are difficult to identify. The aim of this work is to extend the framework of GMR to quantify the reconnection process in this case. 

The paper is structured as follows. In Sec. II, we review the theory of GMR and introduce the relevant mathematical tools. Section III contrasts the manner in which new connections are created for single and multiple reconnection regions. Sections IV and V recap the derivation of the reconnection rate for an isolated region and then derive expressions for the reconnection rate in fragmented current layers. In particular, we show that as with reconnection involving null points a total and a net rate may be defined. The interpretation of each is then discussed. Section VI demonstrates the developed theory for two simple kinematic examples. Finally, Sec. VII summarises the new results and presents our conclusions.

\begin{figure}
\centering
\includegraphics[width=0.4\textwidth]{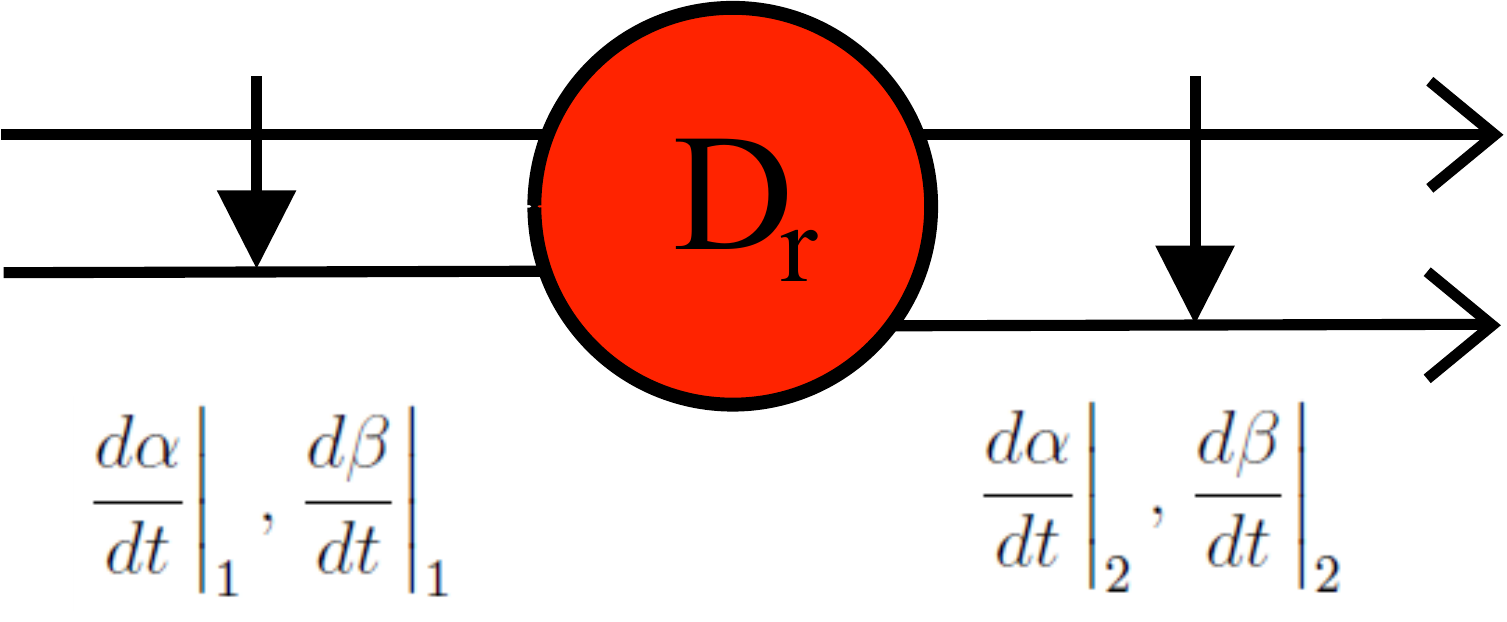}
\caption{Schematic of the difference in the evolutions of $\alpha$ and $\beta$ seen by plasma elements on either side of the non-ideal region, $D_{r}$.}
\label{fig:cart1}
\end{figure}

\begin{figure*}
\centering
\includegraphics[width=0.7\textwidth]{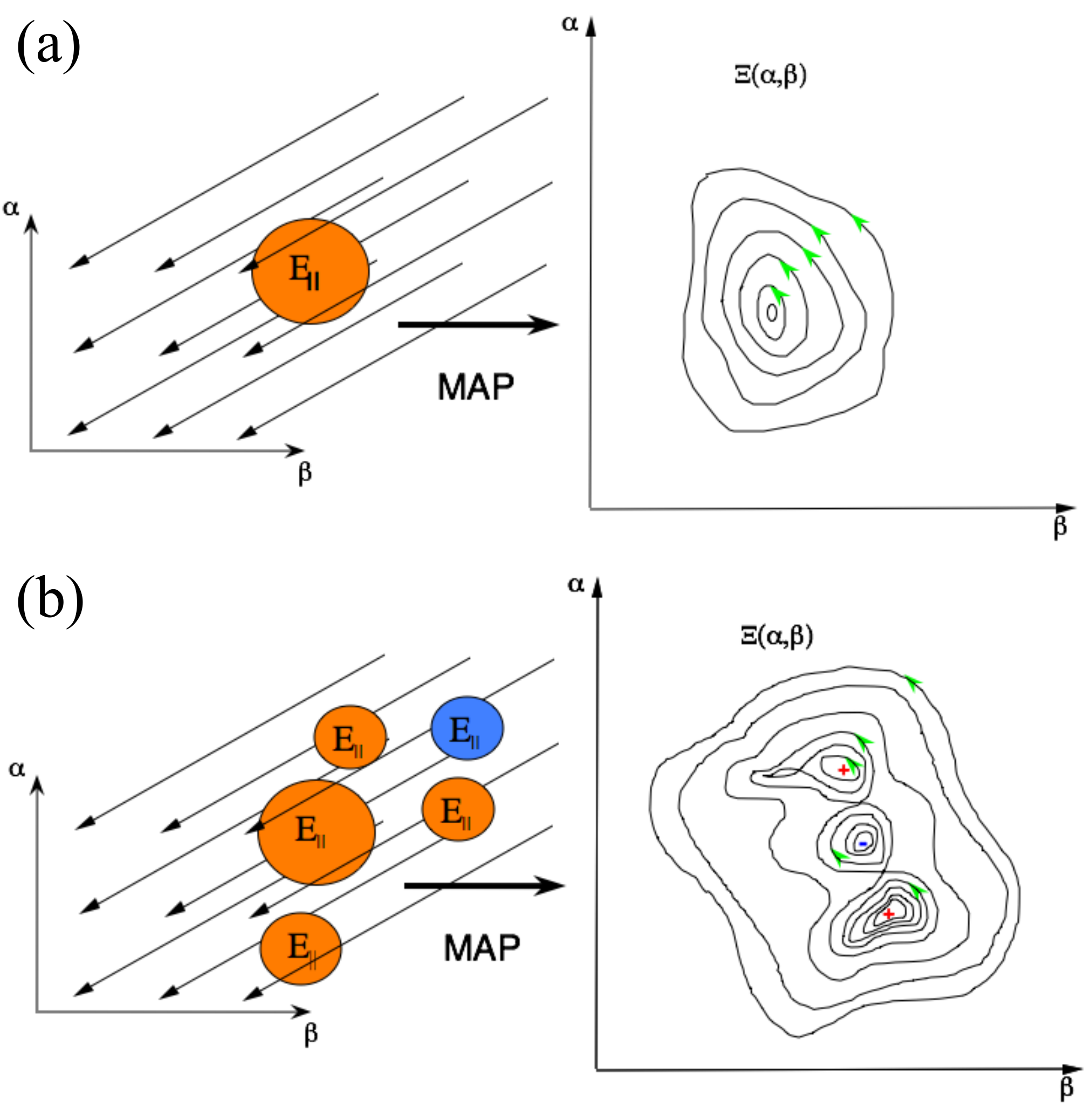}
\caption{ Schematic showing the quasi-potential $\Xi$ mapped into flux coordinate space. (a) A single non-ideal region has one maxima of $\Xi$. (b) Multiple non-ideal regions with multiple maxima and minima in $\Xi$. Arrows show the direction along which new connections form.}
\label{fig:mapping}
\end{figure*}

\section{Overview of General Magnetic Reconnection}
GMR is most readily developed within the framework of Euler potentials \cite{Stern1970}. A pair of Euler potentials ($\alpha$ and $\beta$ say) are scalar functions which locally describe regions of non-vanishing magnetic field through the relation
\begin{align}
\mathbf{B} &=\boldsymbol{\nabla} \alpha\times \boldsymbol{\nabla}\beta. 
\label{bfield}
\end{align}
As long as field lines are simply connected and only enter and leave through the boundaries of the region of interest once, $\alpha$ and $\beta$ are single valued and can be used to label individual field lines. $\alpha$ and $\beta$ are also flux coordinates and are related to the magnetic flux through a given surface via
\begin{equation}
\Phi = \int \int d\alpha \, d\beta.
\label{eulerflux}
\end{equation}
Coupled with an arc length ($s$) satisfying $(\mathbf{B}\cdot \boldsymbol{\nabla}) s=B$, any position within the volume of interest can be expressed in $(\alpha,\beta,s)$ space. Within this formulation the electric field can be expressed as
\begin{equation}
\mathbf{E}=-\frac{\partial \alpha}{\partial t} \boldsymbol{\nabla} \beta + \frac{\partial \beta}{\partial t} \boldsymbol{\nabla} \alpha  -  \boldsymbol{\nabla} \psi,
\label{Efield}
\end{equation}
where the quasi-potential $\psi$ (so named as it contains a time varying component) is related to the electrostatic potential $\phi$ via
\begin{equation}
\psi = \phi +\alpha\frac{\partial \beta}{\partial t},
\label{phi}
\end{equation}
when the magnetic vector potential is assumed to take the form $\mathbf{A} = \alpha \boldsymbol{\nabla} \beta$. See \citet{Hesse1988} for a discussion of the dependance of GMR on the choice of gauge taken for $\mathbf{A}$.

For maximum applicability a general form of Ohm's law is assumed where the contributing non-ideal terms are grouped together into a single vector $\mathbf{R}$ such that
\begin{equation}
\mathbf{E}+ \mathbf{v}\times\mathbf{B} = \mathbf{R},
\label{ohms}
\end{equation}
where $\mathbf{R}$ is assumed to be localized within a small region inside the domain of interest. By expanding $\mathbf{R}$ in covariant form and inserting it into Eqn. (\ref{ohms}) along with Eqns. (\ref{bfield}), (\ref{Efield}) and (\ref{phi})  eventually leads to an expression giving the relative difference between the evolutions of $\alpha$ and $\beta$ that are locally ``seen" by plasma elements on either side of the non-ideal region \cite{Hesse2005,Hesse1988}
\begin{align}
 \left. \frac{d\alpha}{dt}\right|_{2} - \left. \frac{d\alpha}{dt}\right|_{1} &= -\frac{\partial \Xi}{\partial \beta},
 \label{evos1}\\
 \left. \frac{d\beta}{dt}\right|_{2} - \left. \frac{d\beta}{dt}\right|_{1} &= \frac{\partial \Xi}{\partial \alpha},
 \label{evos}
\end{align}
where $\Xi$ is given by
\begin{align}
\Xi(\alpha,\beta) &= -\int_{\alpha,\beta}{E_{\|}ds},\\
&= \psi_{2}-\psi_{1}.
\end{align}
$\psi_{1}$ and $\psi_{2}$ are the  quasi-potential functions on either side of the non-ideal region. Eqns. (\ref{evos1}) and (\ref{evos}) show that plasma elements initially on the same field line threading a localized non-ideal region, $D_{r}$ measure a different evolution of $\alpha$ and $\beta$ and so are not connected by the same field line at a later time. A sketch of this idea is shown in Fig. \ref{fig:cart1}. The power of Eqns. (\ref{evos1}) and (\ref{evos}) is that by considering only the relative difference in the evolutions of plasma elements, ideal flow components are removed, leaving only the components resulting in changes of field line connectivity and thus reconnection. If there is no variation in $\Xi$ in $\alpha\beta$-space then the evolutions of plasma elements are the same on both sides of $D_{r}$. In this case plasma elements which begin on the same field line (and so initially have the same value of $\alpha$ and $\beta$) are subject to the same change in $\alpha$ and $\beta$ and so will be found on the same field line at a later time. Therefore, a necessary and sufficient condition for reconnection is that \cite{Hesse2005,Hesse1993} 
\begin{equation}
\boldsymbol{\nabla}_{\alpha,\beta}\Xi(\alpha,\beta) \neq 0,
\end{equation}
i.e. that there be gradients in $\Xi$ from one field line to another.

\begin{figure*}
\centering
\includegraphics[width=0.9\textwidth]{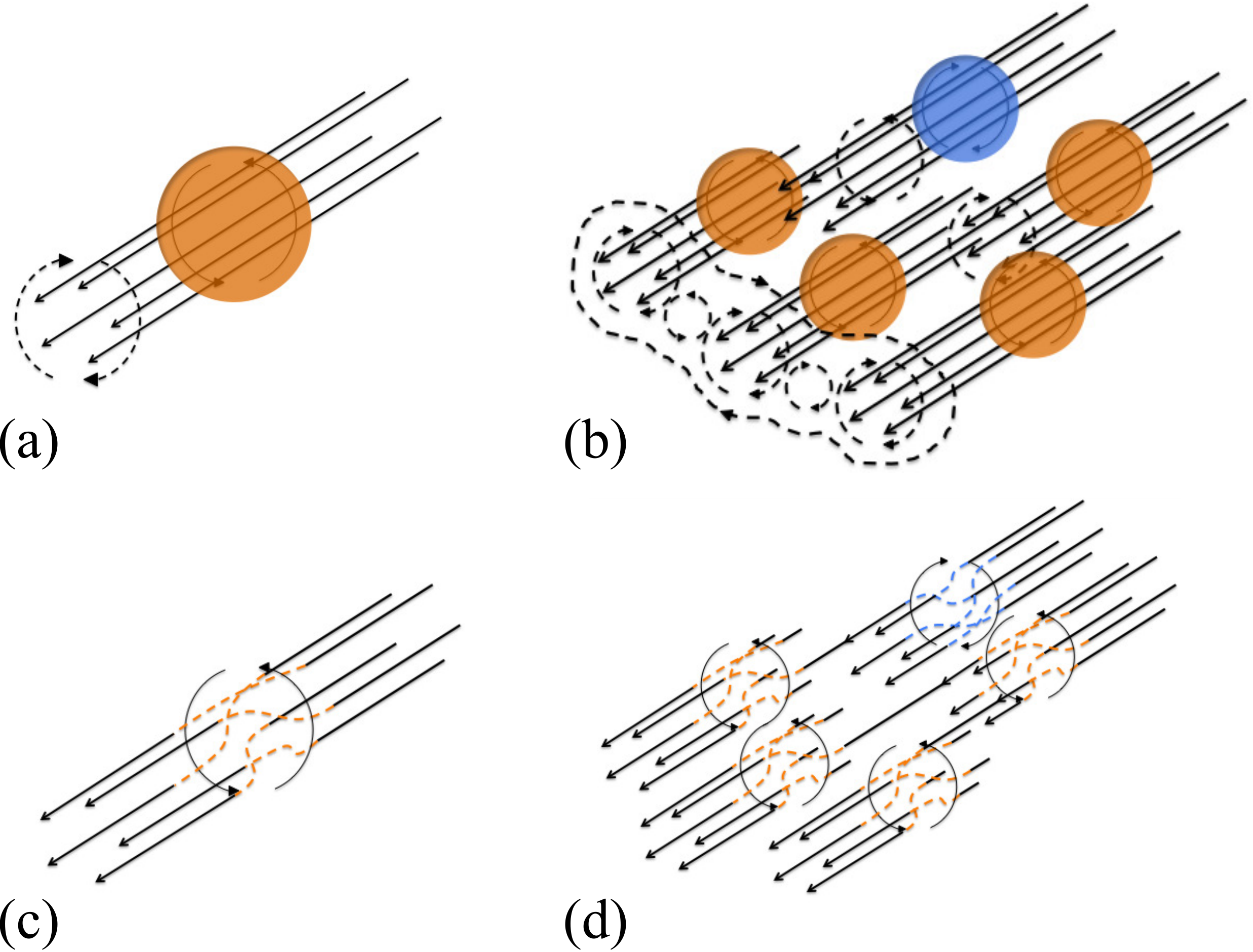}
\caption{The two extremes of connectivity change. Steady state reconnection for a single non-ideal region (a) and multiple non-ideal regions (b).  Purely time dependent reconnection for a single non-ideal region (c) and multiple regions (d). Orange denotes non-ideal regions where $\psi_{2}-\psi_{1}$ is positive and blue where it is negative. Solid arrows show the direction in which new connection form, whilst dashed arrows show induced plasma flows.}
\label{fig:connection}
\end{figure*}

\section{The Nature of the Connectivity Change}
\label{sec:nature}
To understand the nature of any resulting connectivity change for a given problem it is useful to map the problem from 3D real space into flux coordinate ($\alpha\beta$) space. We will work in this space repeatedly throughout the rest of the paper. 

When the reconnection region is assumed to be localized within a single isolated region of $E_{\|}$ the contours of $\Xi$ in flux coordinate space form closed loops. The Hamiltonian nature of Eqns. (\ref{evos1}) and (\ref{evos}) dictates that new connections be formed tangential to the contours of $\Xi$. Thus, when $\Xi$ has only a single extrema these new connections will form in a circular manner. Figure \ref{fig:mapping}(a) shows a sketch of this concept where the green arrows indicate the direction along which new connections form. 

However, when a single reconnection region has an inhomogeneous $E_{\|}$ or multiple reconnection regions exist within the region of interest then the mapping of $\Xi$ in flux coordinate space contains multiple maxima and minima, Fig.  \ref{fig:mapping}(b). In general we restrict ourselves here to scenarios where the multiple $E_{\|}$ regions still only make up a small fraction of the volume under consideration. This means that $\Xi$ approaches zero outside of a flux tube encircling the multiple reconnection sites. The new connections which form now do so along multiple closed loops embedded within a larger scale set of loops,  Fig.  \ref{fig:mapping}(b) (right panel).

The way that this connection change is achieved depends upon the global constraints of the system under consideration. In general the formation of new connections along these loops is a weighted  combination of two extremes:  {\it{steady state}} and purely {\it{time dependent}} connection change \cite{Hesse2005}. It is instructive to consider each in turn.

In steady state the electric field is potential and the magnetic field remains fixed in time. Considering again the case when $\Xi$ has a single extrema, let us then assume that $\mathbf{E}=0$ on one side of the non-ideal region. The only way that new connections can form in the manner shown in Fig. \ref{fig:mapping}(a), whilst also maintaining $\partial \mathbf{B}/\partial t=0$ is by inducing a circular plasma flow of the form shown in Fig. \ref{fig:connection}(a). Ideal flows may be superposed on both sides, however the connection change of the magnetic field within this ideal transporting flow will remain the same. \citet{Hornig2003} considered one such example of this scenario. 

Extending this concept to multiple reconnection regions, each individual non-ideal region will behave locally like the single reconnection region shown in Fig. \ref{fig:connection}(a). The key difference is that now a subset of field lines thread through multiple reconnection regions. Thus, circular plasma flows are induced on field lines leaving a reconnection region which then feed into other secondary regions further along the same field line. Each secondary region superposes a circular plasma flow on to the flow pattern associated with the field lines which thread into it. In some cases this will enhance the induced flow at the exit of the patchy reconnection volume. In others it will act to reduce it. Figure \ref{fig:connection}(b) shows a conceptual sketch of this idea. Thus, steady state patchy reconnection within a localized volume gives rise to an induced localized rotating flow with multiple internal vortices on field lines threading out of the reconnection volume. As with the single reconnection region any background ideal plasma flow may be superposed on to this non-ideal flow.

In the opposite extreme of purely time dependent reconnection the electric field is assumed to be zero on both sides of the non-ideal region. This is particularly relevant to the Solar Corona, e.g. \citet{PriestForbes2000}. In this case new connections can only be formed by a time dependent change in the magnetic field within $D_{r}$. The circular nature of this connection change in situations with a single extrema in $\Xi$ implies that helical magnetic fields are formed in the process, Fig. \ref{fig:connection}(c). When the volume under consideration contains several reconnection sites, each helical region of field may contain a subset of field lines which threads into other helical reconnection regions. Fig. \ref{fig:connection}(d) depicts this idea. This shows that patchy time dependent reconnection can generate (or relax) braided magnetic fields which are thought to be important in the context of coronal heating \cite{Parker1972,Wilmot-Smith2011}. 

In any given 3D reconnection scenario a combination of both manners of connection change are likely to occur.

\begin{figure*}
\centering
\includegraphics[width=0.9\textwidth]{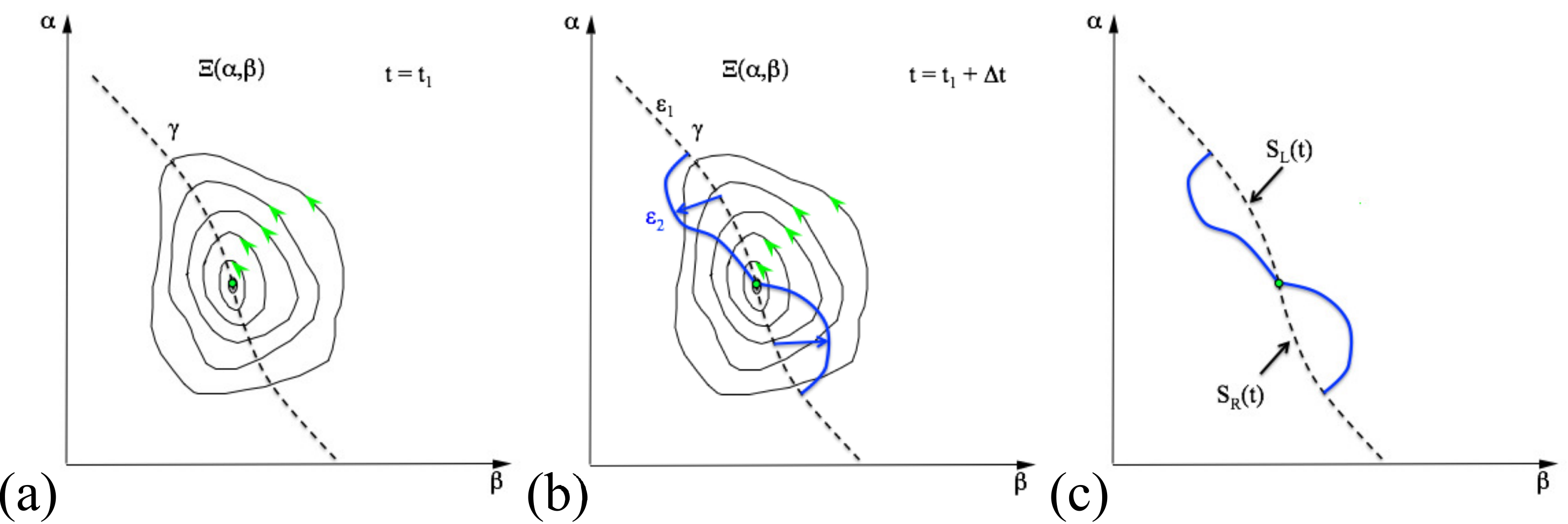}
\caption{The evolution of a flux surface $\gamma$ traversing the non-ideal region and passing through the maxima of $\Xi$ (viewed in flux coordinate space). (a) $\gamma$ (dashed line) over plotting contours of $\Xi$ for a single non-ideal region at $t=t_{1}$. The green point denotes the maxima of $\Xi$ whilst the arrows show the direction of connection change. (b) the evolution of the $\gamma$ surface either side of a single non-ideal region at some later time, $t=t_{1}+\Delta t$. (c) the associated swept out areas of flux ($S_{L}(t)$ and $S_{R}(t)$).}
\label{fig:gamma}
\end{figure*}

\section{Quantifying Reconnection for a Single Reconnection Region}
\label{quant1}
If a given magnetic field contains separatrix surfaces and separator lines then these topological structures can be used as a reference against which the rate of flux transfer may be measured. For instance, when reconnection occurs along a separator the rate of reconnection is simply given by the integral of $E_{\|}$ along the separator line \cite{Vasyliunas1975}. However, in the absence of such structures a more general theory is required. \citet{Hesse2005} developed such a theory for an isolated single reconnection region, $D_{r}$, extending those of previous works \cite{Hesse1993,Hesse1988}. Using a similar approach we now reproduce their results before generalizing the theory to quantify reconnection with multiple reconnection sites. 

Without an obvious reference surface against which to measure \citet{Hesse2005} considered an arbitrary flux surface (i.e. a surface comprised of magnetic field lines) which intersects the single region of parallel electric field and contains the field line along which the integral of $E_{\|}$ is maximal. When mapped into $\alpha\beta$-space this surface appears as a line, which they call the $\gamma$ line. Figure \ref{fig:gamma}(a) shows a sketch of this concept, where the contours depict the quasi-potential $\Xi$.

Now, this flux surface is comprised of field lines embedded in the ideal regions on either side of $D_{r}$. Generally, in each ideal region the evolution is comprised of a background ideal transporting component (which by definition is the same on both sides) and a non-ideal reconnecting component. Without loss of generality we now focus on the non-ideal component by fixing the evolution of field lines threading {\it{into}} the non-ideal region to zero, i.e. $\boldsymbol{\epsilon}_{1}=(d\alpha/dt|_{1},d\beta/dt|_{1})=(0,0)$. This is equivalent to using a coordinate system which moves with the plasma on field lines entering the non-ideal region, allowing the connection change to be entirely characterized by the evolution of the field lines threading {\it{out of}} the non-ideal region which evolve according to $\boldsymbol{\epsilon}_{2}=(d\alpha/dt|_{2},d\beta/dt|_{2})$.

If $\gamma$ is then defined at some arbitrary time ($t=t_{1}$), then at some later time ($t=t_{1}+\Delta t$) the differing evolution on either side of the non-ideal region splits $\gamma$ into two new flux surfaces. In $\alpha\beta$ space these appear as two lines, shown in solid blue and dashed black in Fig. \ref{fig:gamma}(b). Note that as $\boldsymbol{\epsilon}_{1}=(0,0)$ one of these lines is coincident with the original $\gamma$ line. The two new surfaces overlap at the edge of the non-ideal region and at $\Xi_{max}$ (where $\nabla_{\alpha,\beta} \Xi = 0$) since at these places $\boldsymbol{\epsilon}_{1}=\boldsymbol{\epsilon}_{2}=(0,0)$.

The magnetic flux reconnected up to this time is simply given by the flux bounded within one of the two flux tubes formed by these two new flux surfaces, denoted $S_{L}(t)$ and $S_{R}(t)$ (Fig. \ref{fig:gamma}(c)). Each flux tube must have the same cross sectional area 
\begin{equation}
S(t)=S_{L}(t)=S_{R}(t),
\end{equation}
due to the rotational nature of the connection change. In flux coordinate space this area is equal to the magnetic flux within each flux tube, recall the nature of the Euler potentials (Eqn. (\ref{eulerflux})). The rate of reconnection is then defined to be the rate at which $S(t)$ (representing either $S_{L}(t)$ or $S_{R}(t)$) grows at $t=t_{1}$, 
\begin{equation}
\frac{d\Phi}{dt} = \lim_{t\to t_{1}} \left\{ \frac{d}{dt}\int_{S}{d\alpha d\beta} \right\} = \lim_{t\to t_{1}} \left\{ \oint_{\partial S}{\boldsymbol{\epsilon}\cdot \mathbf{n} \,ds } \right \},
\label{recon1}
\end{equation}
where $\boldsymbol{\epsilon} = \boldsymbol{\epsilon}_{1}$ on one side of $S$ and $\boldsymbol{\epsilon} = \boldsymbol{\epsilon}_{2}$ on the other. $\mathbf{n}$ is the outward normal of the boundary $\partial S$. As $t\to t_{1}$ the boundary of $S$ collapses to become the section of the $\gamma$ line on one side of the peak in $\Xi$, referred to hereafter as $\gamma_{1}$. The integral around the boundary of $S$ at $t=t_{1}$ then becomes the superposition of integrals along $\gamma_{1}$, i.e.
\begin{align}
\lim_{t\to t_{1}} \left\{ \oint_{\partial S}{\boldsymbol{\epsilon}\cdot \mathbf{n} \,ds } \right\}= & \int_{\gamma_{1}}{\boldsymbol{\epsilon}_{2}\cdot \mathbf{n} \, dl} - \int_{\gamma_{1}}{\boldsymbol{\epsilon}_{1}\cdot \mathbf{n} \, dl}, \nonumber\\
=&\int_{\gamma_{1}}{\boldsymbol{\epsilon}_{2}\cdot \mathbf{n} \, dl} ,
\label{recon2}
\end{align}
where $l$ is the arc length along $\gamma_{1}$. In coordinate space the local normal to $\gamma_{1}$ is given by 
\begin{equation}
\mathbf{n} = \left( -\frac{\partial \beta}{\partial l}, \frac{\partial \alpha}{\partial l} \right),
\label{normal}
\end{equation} 
whilst Eqns. (\ref{evos1}) and (\ref{evos}) give that
\begin{align}
\boldsymbol{\epsilon}_{2} - \boldsymbol{\epsilon}_{1} & = \boldsymbol{\epsilon}_{2}, \nonumber \\
& = \left(-\frac{\partial \Xi}{\partial \beta}, \frac{\partial \Xi}{\partial \alpha}\right).
\label{diffevo}
\end{align}
Combining Eqns. (\ref{recon1}), (\ref{recon2}), (\ref{normal}) and (\ref{diffevo}) then gives the reconnection rate as
\begin{align}
\frac{d\Phi}{dt} &=  \int_{\gamma_{1}}{ \left(\frac{\partial \Xi}{\partial \beta}\frac{\partial \beta}{\partial l} + \frac{\partial \Xi}{\partial \alpha}\frac{\partial \alpha}{\partial l}\right)} \,dl, \nonumber \\
&= \int_{\gamma_{1}}{\frac{d\Xi}{dl}\, dl} = -\Xi_{max}, \\
&= \left(\int_{\alpha,\beta}{E_{\|}ds}\right)_{max}
\label{rrate}
\end{align} 
Thus, for an isolated region of $E_{\|}$ with a single maximum of $\Xi$ the reconnection rate is given by the value of this maximum. This can be interpreted as the rate at which flux is transferred in one direction across any arbitrarily defined flux surface $\gamma$ which intersects the non-ideal region and includes the field line upon which the maximum of $\Xi$ occurs.

\begin{figure*}
\centering
\includegraphics[width=0.9\textwidth]{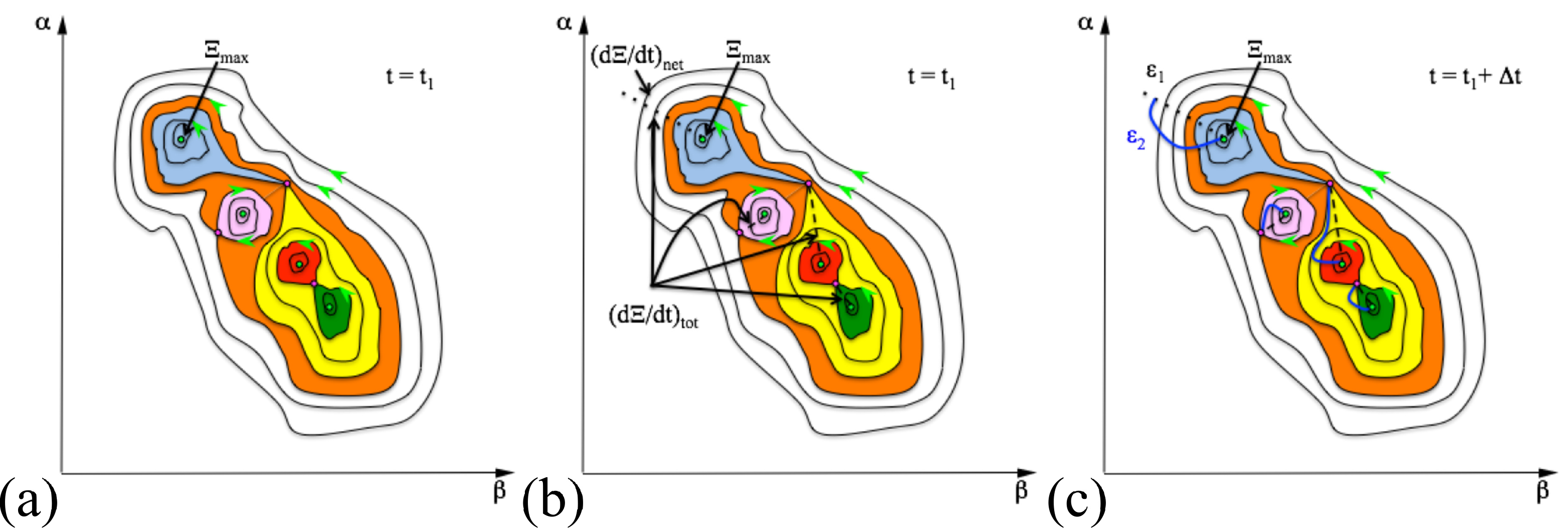}
\caption{The evolution of bounded flux surfaces in a field with multiple reconnection regions (viewed in flux coordinate space). (a) Differently colored regions indicate the different topological regions associated with the connection change. Green circles denote maxima/minima and pink circles saddle points of $\Xi$. (b) dotted line: a flux surface bounded by the the field line along which $\Xi$ is maximum and another in the nearby ideal region, used to calculate $(d\Phi/dt)_{net}$ (see text). Dashed lines: flux surfaces bounded by field lines that have local extrema in $\Xi$ and field lines at the nearest associated saddle point, used to calculate $(d\Phi/dt)_{tot}$ (see text). (c) the associated swept out areas of flux for each bounded flux surface.}
\label{fig:arc}
\end{figure*}

\begin{figure*}
\centering
\includegraphics[width=0.95\textwidth]{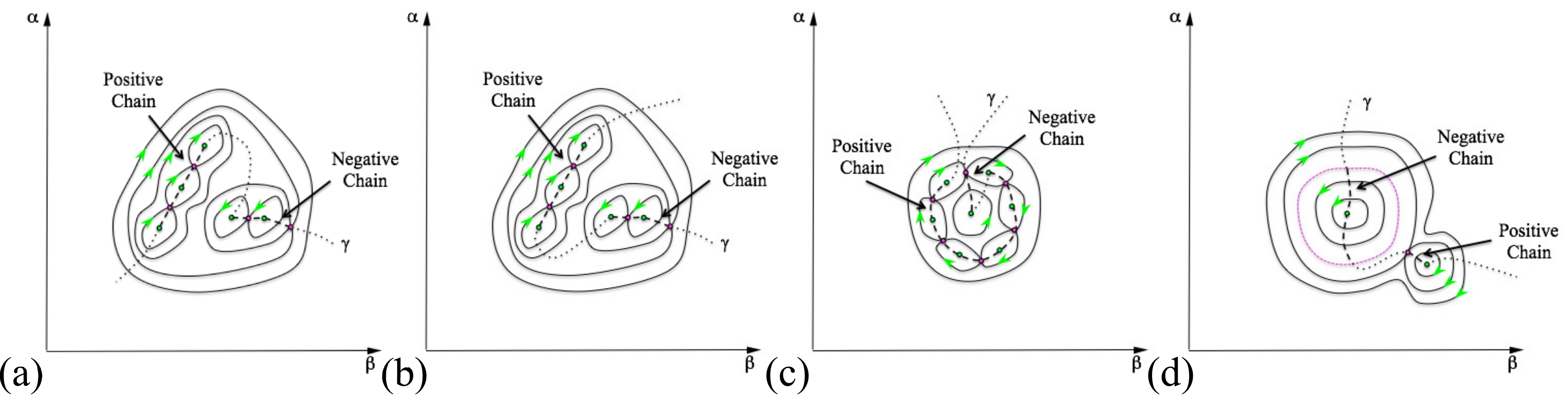}
\caption{Examples of flux surfaces ($\gamma$) chosen so that the extrema and saddle points of $\Xi$ are connected in a certain way (see text for details). Green and pink dots denote extrema and saddle points respectively. Arrows depict the direction of connection change. Dashed lines show sections of $\gamma$ forming chains of maxima and minima. Dotted lines show sections of $\gamma$ connecting these chains with each other or the background ideal field. (a)-(b) the same $\Xi$ profile with two different choices for $\gamma$. (c) a choice for $\gamma$ in a more complex field. (d) a choice for $\gamma$ in a degenerate $\Xi$ field with a circular maxima (pink dashed line).}
\label{fig:chains}
\end{figure*}

\begin{figure*}
\centering
\includegraphics[width=0.7\textwidth]{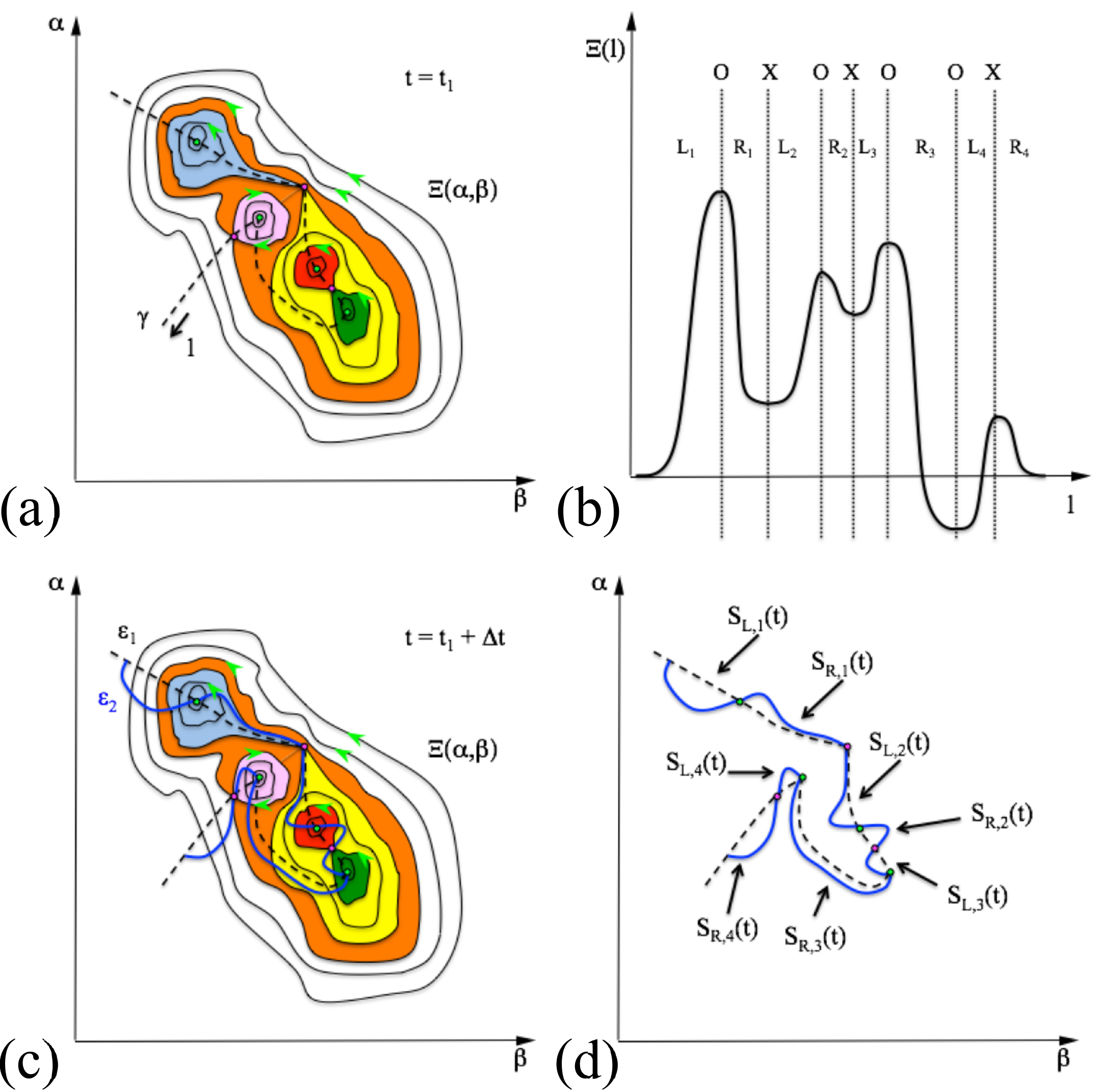}
\caption{Interpretation of reconnection rate in terms of a global flux surface. (a) Dashed line shows a flux surface ($\gamma$) which traverses all of the non-ideal regions and includes the field lines at which the extrema and saddle points of $\Xi$ are found. (b) variation of $\Xi$ as a function  of arc length $l$ along the $\gamma$ line in flux coordinate space. $L_{i}$ and $R_{i}$ denote the direction of flux transfer across $\gamma$ (where $i \in [1,4]$) and the $X$'s and $O$'s show the positions of the saddle points and extrema of $\Xi$ respectively. (c) the evolution of the $\gamma$ surface either side of the multiple non-ideal regions and (d) the associated swept out areas of flux, $S_{L,i}(t)$ and $S_{R,i}(t)$ corresponding to the different sections of flux transfer across $\gamma$ ($L_{i}$ and $R_{i}$).}
\label{fig:area}
\end{figure*}

\section{Generalization to Multiple Reconnection Sites}
When there are multiple reconnection sites or inhomogeneity of $E_{\|}$ within a single site there is likely to be multiple peaks in $\Xi$. We now aim to develop expressions which quantify the rate of reconnection in this case and explain their interpretations. 

\subsection{Expressions for Reconnection rate}
As discussed in Sec. \ref{sec:nature}, when there are multiple peaks in $\Xi$ new connections are formed along multiple embedded closed paths in the $\alpha\beta$ plane. Near positive extrema (peaks) of $\Xi$ the direction that this new connection formation takes is clockwise, whereas for negative extrema (troughs) it is anti-clockwise, Fig. \ref{fig:mapping}(b). The places where there is no connection change occur where $\boldsymbol{\nabla}_{\alpha,\beta}\Xi(\alpha,\beta) = 0$. These correspond to the field lines not threading into any non-ideal region (the neighboring ideal field) and special field lines along which the net difference in connection change along their length is zero, i.e. field lines along which the induced connection change from multiple reconnection sites cancels out. These special field lines sit at the critical points (``X-points" and ``O-points") of the divergence free field defining the direction of new connection formation:
\begin{equation}
\boldsymbol{\epsilon}_{2}= \left(-\frac{\partial \Xi}{\partial \beta}, \frac{ \partial \Xi}{\partial \alpha} \right).
\end{equation} 
In terms of the quasi-potential the ``O-points'' correspond to peaks and troughs of $\Xi$, whereas the ``X-points'' occur at saddle points. Figure \ref{fig:arc}(a) shows a sketch of this idea, where the green and pink circles show the position of the ``O-points''  and ``X-points''  respectively. The key idea here is that just like X-points divide up two dimensional magnetic fields into distinct topological regions, so also the rotational formation of new connections described by $\boldsymbol{\epsilon}_{2}$ is partitioned into localized rotational regions (with``O-points" at their centers) by a series of ``X-points''. The different topological regions of the  $\boldsymbol{\epsilon}_{2}$ field are shown in different colors in Fig. \ref{fig:arc} to better illustrate them.

\begin{figure}
\centering
\includegraphics[width=0.33\textwidth]{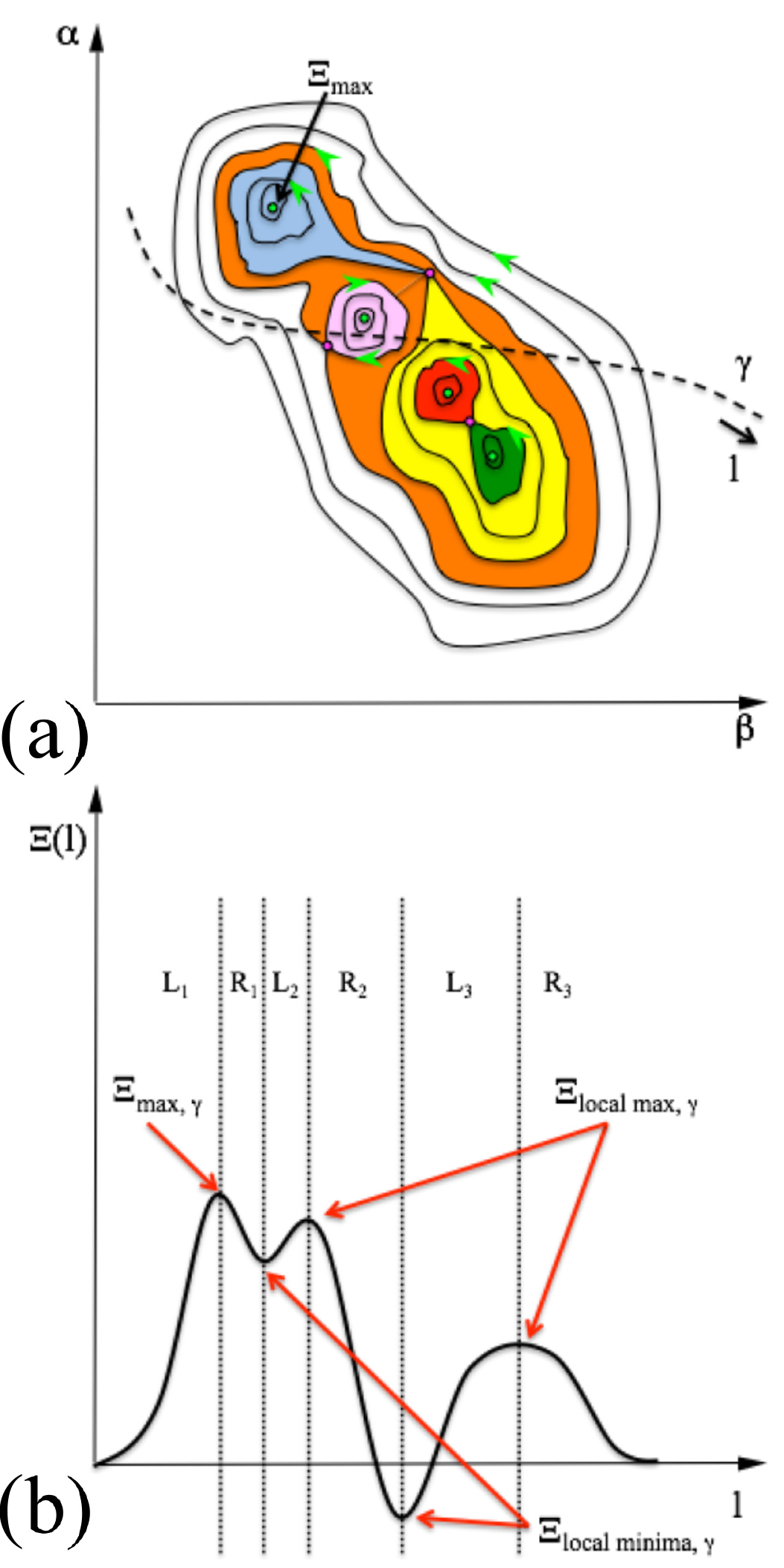}
\caption{Reconnection of flux across a specific flux surface. (a) dashed line depicts $\gamma$ in flux coordinate space. The contours depict $\Xi$ and the colors denote the topological regions associated with $\boldsymbol{\epsilon}_{2}$. (b) $\Xi$ as a function of arc length $l$ along $\gamma$. $L_{i}$ and $R_{i}$ (where $i\in [1,3]$) denote the direction of flux transfer across $\gamma$. }
\label{fig:gen_gamma}
\end{figure}

To construct expressions for the reconnection rate we begin in the same manner as Sec. \ref{quant1} and consider a flux surface bounded by the field line along which $\Xi_{max}$ occurs and some other field line in the nearby ideal region. Figure \ref{fig:arc}(b) shows this surface as a dotted line in flux coordinate space. Now, the only  topological regions of $\boldsymbol{\epsilon}_{2}$ that are straddled by this surface are the region within which $\Xi_{max}$ is situated (light blue) and the regions within which the surrounding outer loops of connect change occur (white and orange regions). The connection change within these regions gives the net rate at which new connections form in a rotational manner around the field line on which $\Xi_{max}$ occurs. The same analysis as Sec. \ref{quant1} may then be applied to this flux surface, quantifying this net rate as
\begin{equation}
\left(\frac{d\Phi}{dt}\right)_{net} = -\Xi_{max} = \left(\int_{\alpha,\beta}{E_{\|}ds}\right)_{max}.
\label{rrnet}
\end{equation}
Thus, the maximum of the integral of $E_{\|}$ across a fragmented reconnection region measures a net rate of rotational connection change and neglects the connection change associated with other extrema of $\Xi$. 

To quantify the true rate of reconnection requires that the connection change associated with each of these other extrema also be taken account of. This can be achieved by considering a flux surface for each additional extrema bounded on one side by the field line at which the extrema occurs and on the other by the field line situated at the nearest saddle point (s.p.) of $\Xi$ (corresponding to each additional ``O-point'' and its nearest ``X-point'' of the $\boldsymbol{\epsilon}_{2}$ field). Figure \ref{fig:arc}(b) illustrates this idea with a series of dashed lines in flux coordinate space. For each of these additional flux surfaces one can also apply the same analysis as Sec. \ref{quant1} to give the rate of connection change across the surface as
\begin{align}
\left(\frac{d\Phi}{dt}\right)_{local}  &=  \int{\frac{d\Xi}{dl}\, dl}, \nonumber\\
&= \Xi_{\text{local extrema}}-\Xi_{\text{nearest s.p.}}.
\end{align}
The total reconnection rate associated with all of the non-ideal regions is then given by the sum of the local connection change occurring around each additional extrema in addition to the net rotational connection change occurring around the largest extrema, i.e.
\begin{align}
\left(\frac{d\Phi}{dt}\right)_{tot} = \left|\Xi_{\text{max}}\right| + \sum_{i} \left| \Xi_{\text{local extrema}, i} - \Xi_{\text{adjacent\, s.p.},i}\right|,
\label{rrtot}
\end{align}
where absolute values have been used to account for when $\Xi$ has both maxima and minima.

The above shows that when reconnection occurs in fragmented reconnection regions the rate of reconnection can be quantified by two different measurements. The first, $(d\Phi/dt)_{tot}$ describes the true rate at which new connections are formed collectively by the fragmented layer. The second, $(d\Phi/dt)_{net}$ gives the net rate of flux transfer associated with the maximal value of $\Xi$ on field lines crossing the volume. Both measurements are equivalent when there is only one peak in $\Xi$.

\subsection{Interpretation in Terms of Flux Transferred Across a Single Large Flux Surface}
\label{sec:fluxsurf}
The above analysis shows that reconnection in fragmented current layers can be considered as representing the rate at which magnetic flux is reconnected across multiple bounded flux surfaces. We now show that provided $\Xi$ is smooth and continuous, $(d\Phi/dt)_{tot}$ and $(d\Phi/dt)_{net}$ can also be interpreted as the total and net rate of flux transfer across a large scale flux surface ($\gamma$) spanning the entire reconnection volume.

From the above analysis one would expect that such a flux surface must contain the field lines along which each of the extrema and saddle points of the $\Xi$ profile occur. However, the order in which each extrema and saddle points are connected by $\gamma$ is crucial. In particular, $\gamma$ must be defined such that extrema of the same type (maxima or minima) are connected via any adjoining saddle points forming chains. The end of a chain of maxima can be connected with the end of a chain of minima if the connection is from the maxima to minima, or saddle point to saddle point. Alternatively, the ends of chains of maxima or minima may instead be connected with the surrounding ideal field. Figure \ref{fig:chains}(a)-(c) shows three examples. In the degenerate case of when a local maxima of $\Xi$ forms a ring, any two points on the ring may be chosen in place of two saddle points, Fig. \ref{fig:chains}(d).

We point out that the selection of this flux surface is not unique and differs depending upon how different chains of maxima and minima are connected. Figure \ref{fig:chains}(a)-(b) illustrates this idea by the differing dotted sections of $\gamma$. 

The dashed line depicted in Fig. \ref{fig:area}(a) shows a flux surface which connects the extrema and saddle points of the previously considered $\Xi$ profile in the way described above way at some arbitrary time ($t=t_{1}$). By choosing $\gamma$ in this way the flux transfer between the critical points of $\Xi$ alternates along $\gamma$. This is shown in Fig. \ref{fig:area}(b) which depicts the variation of $\Xi$ along $\gamma$ as a function of the arc length ($l$). The positions of the extrema are shown with ``O's'' and the saddle points with ``X's'', whilst the direction of flux transfer across $\gamma$ between them is indicated by ``$L_{i}$'' or ``$R_{i}$'', where $i \in [1,4]$.

At some later time ($t=t_{1}+\Delta t$) the differing field evolutions on either side of the multiple non-ideal regions forms a chain of flux tubes with cross sectional areas of $S_{L,i}(t)$ or $S_{R,i}(t)$  associated with $L_{i}$ and $R_{i}$ respectively. Note, that the rotational nature of $\boldsymbol{\epsilon}_{2}$ means that the sum of each set of area elements must be the same, i.e. 
\begin{equation}
S(t) =  \sum_{i} \left\{\int_{S_{L,i}}{d\alpha d\beta}\right\} = \sum_{i} \left\{\int_{S_{R,i}}{d\alpha d\beta}\right\}.
\end{equation}
Now, if we compare the area segments swept out by the series of bounded flux regions discussed earlier (Fig. \ref{fig:arc}(c)) to those generated by this continuous surface (Fig. \ref{fig:area}(b)) we find that they match $S_{L,i}(t)$. This shows that the total rate of reconnection can be interpreted as the rate of growth as $t\to t_{1}$ of the collective area $S(t)$ associated with flux swept in the same direction (all to the left {\it{or}} all to the right) across $\gamma$, i.e.
\begin{equation}
\left(\frac{d\Phi}{dt}\right)_{total}  = \lim_{t\to t_{1}}\left\{\frac{d}{dt} S(t) \right\}.
\end{equation}
A similar conclusion can also be drawn for other choices of $\gamma$ connecting the chains of maxima and minima. 

Similarly, the net rate can be interpreted as the rate of growth as $t\to t_{1}$ of the {\it{difference}} in the areas associated with flux swept in one (or the other) direction on one side of $\Xi_{max}$, i.e.
\begin{equation}
\left(\frac{d\Phi}{dt}\right)_{net}  =  \lim_{t\to t_{1}}\left\{\frac{d}{dt} S_{d}(t) \right\},
\end{equation}
where
\begin{equation}
S_{d}(t) =  \sum_{j} \left\{\int_{S_{L,j}}{d\alpha d\beta}\right\} - \sum_{k} \left\{\int_{S_{R,k}}{d\alpha d\beta}\right\},
\end{equation}
and $j$ and $k$ sum over the area segments formed along the portion of $\gamma$ on one (or other) side of $\Xi_{max}$.

It should be noted that the existence of such a large scale $\gamma$ surface is not necessary for the application of Eqns. (\ref{rrnet}) and (\ref{rrtot}), and indeed if $\Xi$ is sufficiently complex or contains discontinuities such a surface may not be definable. However, we have shown that at least when $\Xi$ is smooth and relatively simple the intuitive idea that the reconnection rate should measure the rate at which flux is reconnected across some large scale flux surface (akin to that of a true separatrix when reconnection occurs between distinct topological regions) still holds.

\subsection{Quantifying Reconnection Across and Arbitrary Flux Surface}
Finally, we now consider the case where rather than wanting to know the true rate of reconnection, one is interested in knowing the rate at which flux is reconnected past a particular flux surface. An example of such a surface would be one associated with an observed flare ribbon on the photosphere. Another would be if the global topology is such that field lines from a separatrix surface or surfaces pass through the domain of interest and one wishes to know the rate of flux transfer between two different topological domains.

Equations (\ref{rrnet}) and (\ref{rrtot}) are easily generalized to this scenario. Consider some arbitrary flux surface spanning a fragmented reconnection region with multiple peaks in $\Xi$, Fig. \ref{fig:gen_gamma}(a). Along the length of $\gamma$ a number of local maxima and minima of $\Xi$ occur. Between each of these local extrema flux is transferred in one or other direction depending upon the sign of the gradient of $\Xi(l)$, Fig. \ref{fig:gen_gamma}(b). In analogy to the previous sections the net rate at which flux is transferred across this surface is given by
\begin{equation}
\left(\frac{d\Phi}{dt}\right)_{net,\gamma} = -\Xi_{max,\gamma} = \left(\int_{\alpha,\beta}{E_{\|}ds}\right)_{max,
\gamma},
\label{rrnetloc}
\end{equation}
where the subscript $\gamma$ denotes measurement of each quantity along the $\gamma$ line. Similarly the total rate of flux transfer across this particular flux surface is given by
\begin{equation}
\left(\frac{d\Phi}{dt}\right)_{tot,\gamma} = \left|\Xi_{\text{max}, \gamma}\right| + \sum_{i} \left| \Xi_{\text{local max}, i} - \Xi_{\text{adjacent min},i}\right|_{\gamma}.
\label{rrtotloc}
\end{equation}
Depending upon the path take by the $\gamma$ line as it crosses $\Xi$ in flux coordinate space the value of $(d\Xi/dt)_{tot,\gamma}$ can be greater or less than the value measured by Eqn. (\ref{rrtot}). For instance if $\gamma$ is chosen so that it crosses $\Xi$ many times, then it would be likely that $(d\Xi/dt)_{tot,\gamma} > (d\Xi/dt)_{tot}$. However, by definition the net rate of transfer will at most be the same as the net rate of rotational connection change around the field line with $\Xi = \Xi_{max}$, so that $(d\Xi/dt)_{net,\gamma} \le (d\Xi/dt)_{net}$.

\begin{table}[ht]
\caption{Model Parameters} 
\centering 
\begin{tabular}{c c c c c c c c} 
\hline\hline 
$i$ & $j_{i}$ & $l_{x,i}$ & $l_{y,i}$ & $l_{z,i}$ & $x_{0,i}$ & $y_{0,i}$ & $z_{0,i}$ \\ [0.5ex] 
\hline 
1 & -0.1 & 0.2 & 0.2 & 0.2 & 0.0 & 0.0 & 0.0\\ 
2 & -0.1 & 0.1 & 0.1 & 0.1 & 0.0 & 0.35 & -1.0\\
3 & -0.1 & 0.1 & 0.1 & 0.1 & 0.0 & -0.35 & -1.0\\[1ex] 
\hline 
\end{tabular}
\label{table:runs} 
\end{table}

\begin{figure}
\centering
\includegraphics[width=0.45\textwidth]{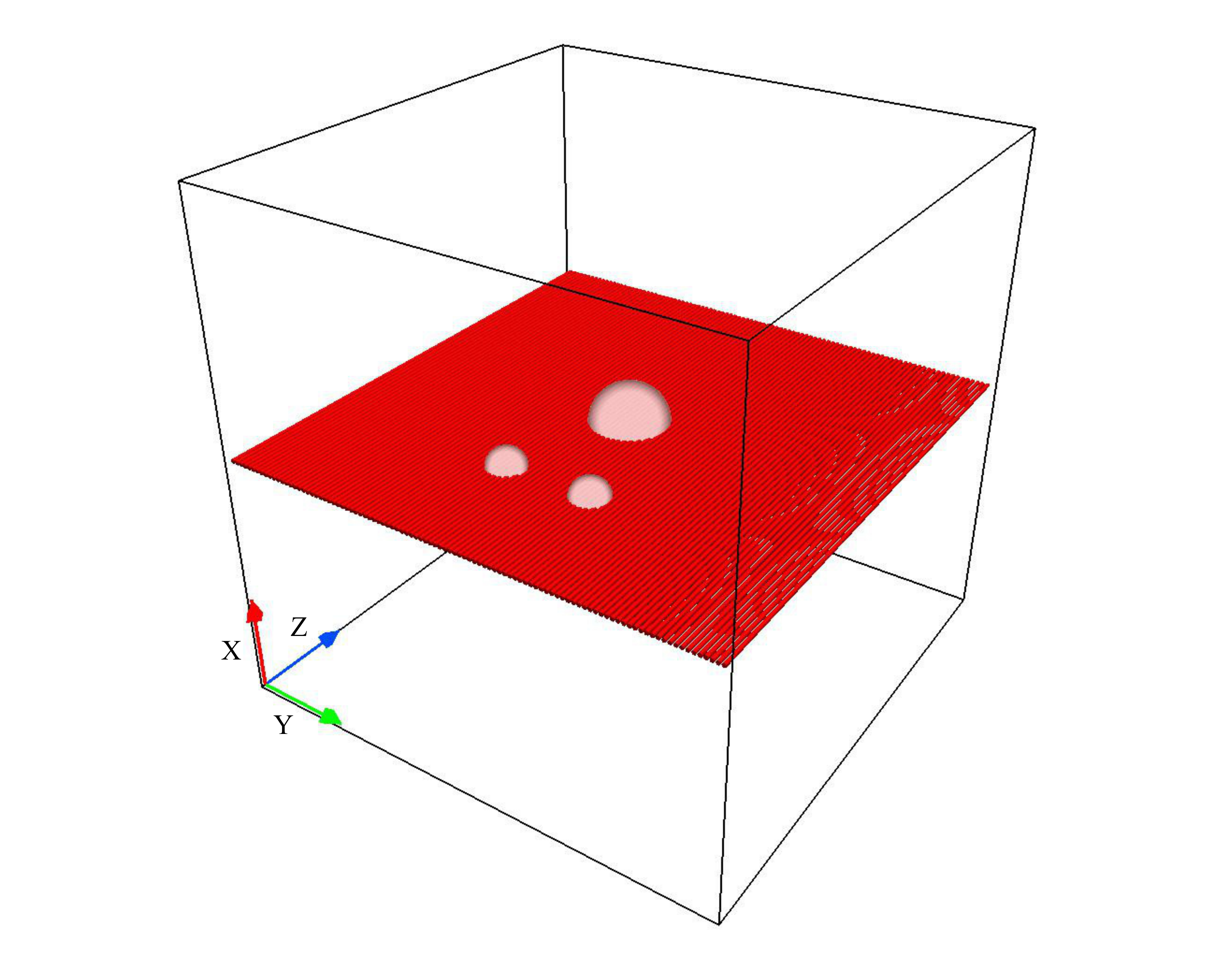}
\caption{Iso-surfaces at $10\%$ of the maximum of $|\mathbf{R}|$, showing the three localized non-ideal regions at $t=0$ in both models. In red are a selection of field lines plotted from footpoints along $(x,z) = (0,2)$. }
\label{fig:vapor_t0}
\end{figure}

\section{Example: Multiple Reconnection Sites in a Straight Field} 
To illustrate the theory we now present two simple kinematic models of an idealized fragmented current layer.

Starting with an initial magnetic field (at $t=0$) of the form
\begin{equation}
\mathbf{B} =  B_{0}\,\hat{\mathbf{z}},
\end{equation}
we assume some non-ideal process occurs to produce multiple non-ideal regions such that
\begin{equation}
\mathbf{R} = \sum_{i=0}^{n}  j_{i} e^{-(x-x_{0,i})^2/l_{x,i}^2 -(y-y_{0,i})^2/l_{y,i}^2 -(z-z_{0,i})^2/l_{z,i}^2}  \,\hat{\mathbf{z}}.
\end{equation}
where ($l_{x,i}, l_{y,i}, l_{z,i}$), ($x_{0,i}, y_{0,i}, z_{0,i}$) and $j_{i}$ control the dimensions, position and the strength respectively of each non-ideal region. We choose three non-ideal regions ($n=3$), one larger central region and two smaller identical offset regions, see Fig. \ref{fig:vapor_t0}. The chosen parameter values are given in Table \ref{table:runs}. Depending upon the constraints placed upon the system, reconnection solutions describing purely time dependent, steady state or a combination of both scenarios can be constructed. In what follows we will consider the two extreme cases and verify in each case the validity of the Eqns. (\ref{rrnet}) and (\ref{rrtot}). 

\begin{figure}
\centering
\includegraphics[width=0.45\textwidth]{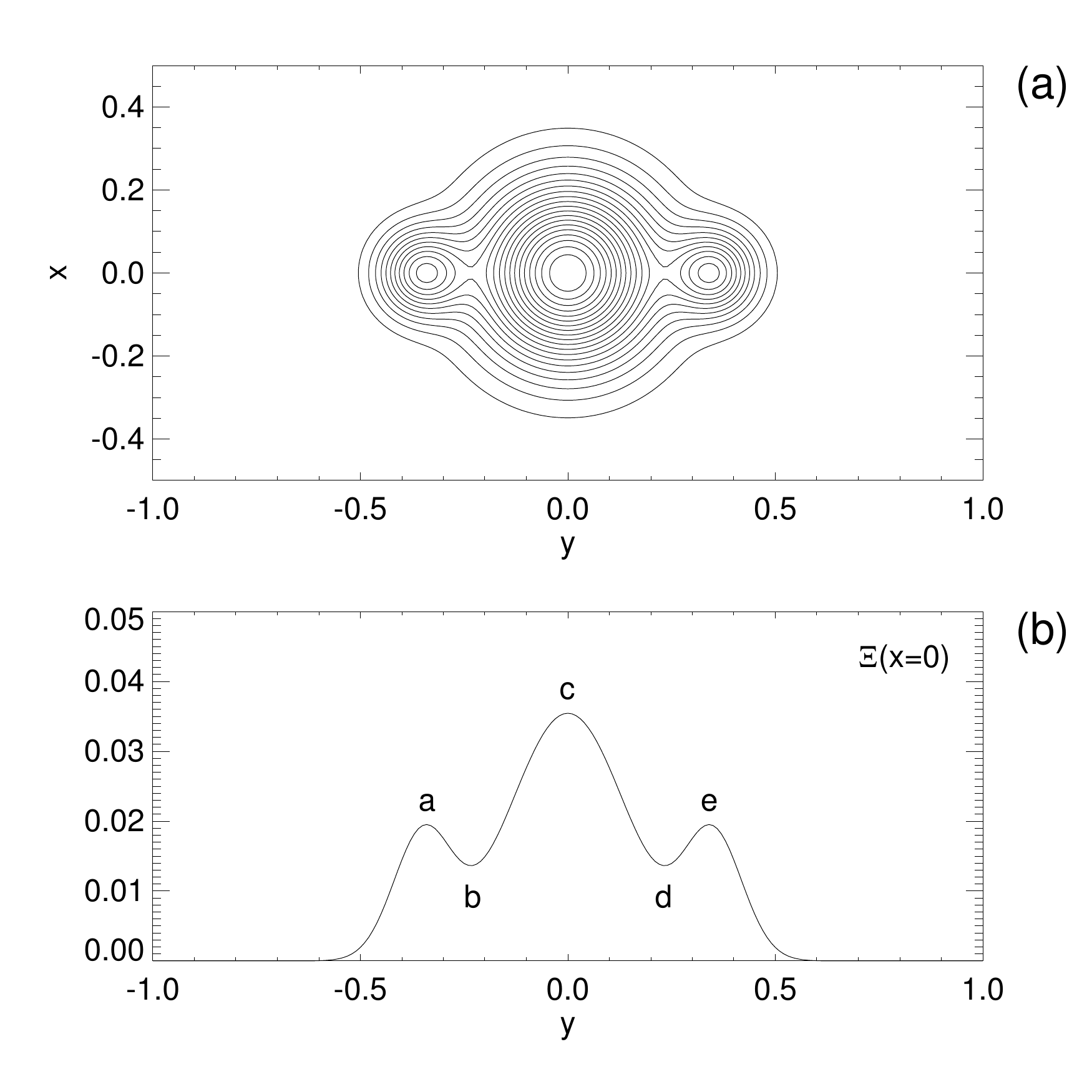}
\caption{(a) contours of $\Xi(x,y)$ mapped on to flux coordinate space. (b) $\Xi$ along the line $x=0$, passing through the five critical points. }
\label{fig:phi_t0}
\end{figure}

\subsection{Time Dependent Reconnection}
In this extreme we impose that the sections of field lines threading into and out of the non-ideal region are held fixed such that the electric field vanishes on each side of the non-ideal region. This is equivalent to assuming that the plasma velocity $\mathbf{v}=0$ everywhere. Ohm's law then gives directly that $\mathbf{E}=\mathbf{R}$, i.e.
\begin{equation}
\mathbf{E} = \sum_{i=0}^{3}  j_{i} e^{-(x-x_{0,i})^2/l_{x,i}^2 -(y-y_{0,i})^2/l_{y,i}^2 -(z-z_{0,i})^2/l_{z,i}^2}  \,\hat{\mathbf{z}}.
\end{equation}
Faraday's law, $\partial \mathbf{B}/\partial t=\nabla \times \mathbf{E}$ then dictates that at later times the magnetic field evolves such that
\begin{equation} 
\mathbf{B} =  B_{0}\,\hat{\mathbf{z}} + \sum_{i=0}^{3}\nabla \times  \mathbf{A}_{i,flux ring},
\end{equation}
where $i$ sums over each non-ideal region and 
\begin{equation}
\mathbf{A}_{i,flux ring}  = t  j_{i} e^{-(x-x_{0,i})^2/l_{x,i}^2 -(y-y_{0,i})^2/l_{y,i}^2 -(z-z_{0,i})^2/l_{z,i}^2}  \,\hat{\mathbf{z}}.
\end{equation}
At $t=0$ the magnetic field is initially straight, but as time progresses each flux ring introduces an ever increasing twist to the field. Note that in this simple example we are only considering small periods in time, $t$. At $t=0$ the straight magnetic field can be described with the two Euler potentials $\alpha = x$ and $\beta =y$. Since each non-ideal region is negligibly strong in the vicinity of the others $\Xi(\alpha,\beta)=\Xi(x,y)$ can be constructed from the superpositions of $\int_{-\infty}^{+\infty}{E_{\|}dl}=\int_{-\infty}^{+\infty}{E_{z}dz}$ across each region giving 
\begin{equation}
\Xi(x,y) = \sqrt{\pi}\sum_{i=0}^{n} \frac{j_{i}}{l_{z,i}} e^{-(x-x_{0,i})^2/l_{x,i}^2 -(y-y_{0,i})^2/l_{y,i}^2} + f(x,y)
\end{equation}
where $f(x,y)$ is an arbitrary function. In what follows we will trace field lines from $z=+2$ to $z=-2$ so for convenience we set $\Xi(x,y)=0$ at $z=+\infty$ to give
\begin{equation}
\Xi(x,y) = -\sqrt{\pi}\sum_{i=0}^{n} \frac{j_{i}}{l_{z,i}} e^{-(x-x_{0,i})^2/l_{x,i}^2 -(y-y_{0,i})^2/l_{y,i}^2}
\end{equation} 
Figure \ref{fig:phi_t0}a shows a contour plot of $\Xi$ at $t=0$ mapped on to the $xy$-plane. The profile contains three distinct peaks (O-points) with two saddle points (X-points) between them. By symmetry the X-points and O-points of $\boldsymbol{\epsilon}_{2}-\boldsymbol{\epsilon}_{1}=\boldsymbol{\epsilon}_{2}$ lie along $x=0$, so we choose this as  our $\gamma$ line. 

The variation of the quasi-potential along this line $\Xi(x=0,y)$ is shown in Fig. \ref{fig:phi_t0}b. The peaks occur at $y=a, c$ and $e$, with the saddle points located at $y=b$ and $d$. Applying Eqn. (\ref{rrtot}) gives the total reconnection rate of this system as
\begin{align}
 \left(\frac{d\Phi}{dt}\right)_{tot} &= \Xi_{\text{max}} + \sum_{i}\left| \Xi_{\text{local extrema};i} - \Xi_{\text{adjacent s.p.};i}\right| \nonumber \\
 	&= \Xi_{c} + \left|\Xi_{a}-\Xi_{b}\right| + \left|\Xi_{e}-\Xi_{d}\right| \nonumber \\
	&\approx 0.04711,
\end{align} 
with a net rate of flux transfer given by
\begin{equation}
 \left(\frac{d\Phi}{dt}\right)_{net}  = \Xi_{max} \approx 0.03545.
 \end{equation}
 In this extreme, these values represent the total and net rate respectively at which magnetic field is generated normal to the $\gamma$ surface collectively by the non-ideal regions.

We now go on to verify these values by comparing them with values obtained numerically from a flux counting procedure, explained below. A large number of field lines were traced from a grid on $z=2$ as far as $z=-2$. At both positions the magnetic field has reached its asymptotic value of $\mathbf{B} =  B_{0}\,\hat{\mathbf{z}}$. This is done for the field at some time, $t=t_{1}$ and some later time $t=t_{1}+\Delta t$. The amount of flux transfer ($\Delta \Phi$) in this period is obtained by comparing the final positions (on $z=-2$) at both times and summing the number of field lines to have crossed the $\gamma$ line, weighted by their area element on the starting grid and the field strength perpendicular to the surface of starting points, i.e.
\begin{equation}
\Delta \Phi = \sum_{N}{B_{0}\Delta x\Delta y},
\end{equation}
where $N$ is the number of field lines under consideration. The rate of reconnection is then estimated as
\begin{equation}
\frac{d\Phi}{dt} \approx \frac{\Delta \Phi}{\Delta t}.
\end{equation}
To obtain $(d\Phi/dt)_{tot}$ all field lines found to have crossed $\gamma$ in $\Delta t$ are counted and the value halved so as not to double count the flux transfer (recall that the connection change is circular and so will cross the $\gamma$ line twice). $(d\Phi/dt)_{net}$ is approximated by counting only the net transfer across a half segment of the $\gamma$ line. 

\begin{figure}
\centering
\includegraphics[width=0.5\textwidth]{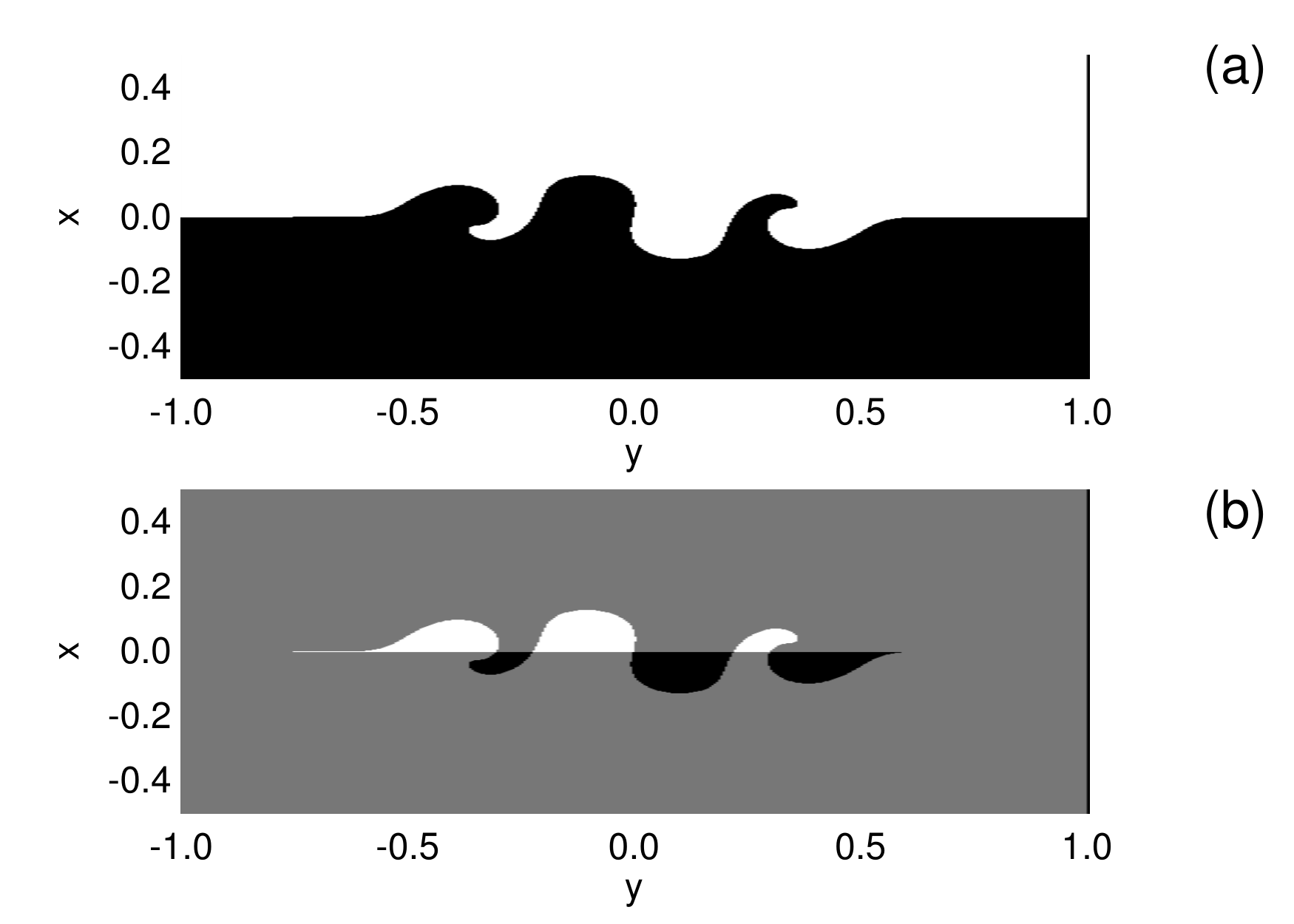}
\caption{(a) connectivity map at $t=1$ for the time dependent model. Black show field lines with starting points below $x=0$ and white those with starting points above. (b) connectivity plot of the field lines to have changed connection between $t=0$ and $t=1$. Black regions have moved from $x<0$ to $x>0$, white have moved from $x>0$ to $x<0$ and grey regions have stayed the same.}
\label{fig:fluxcount}
\end{figure}

\begin{figure}
\centering
\includegraphics[width=0.45\textwidth]{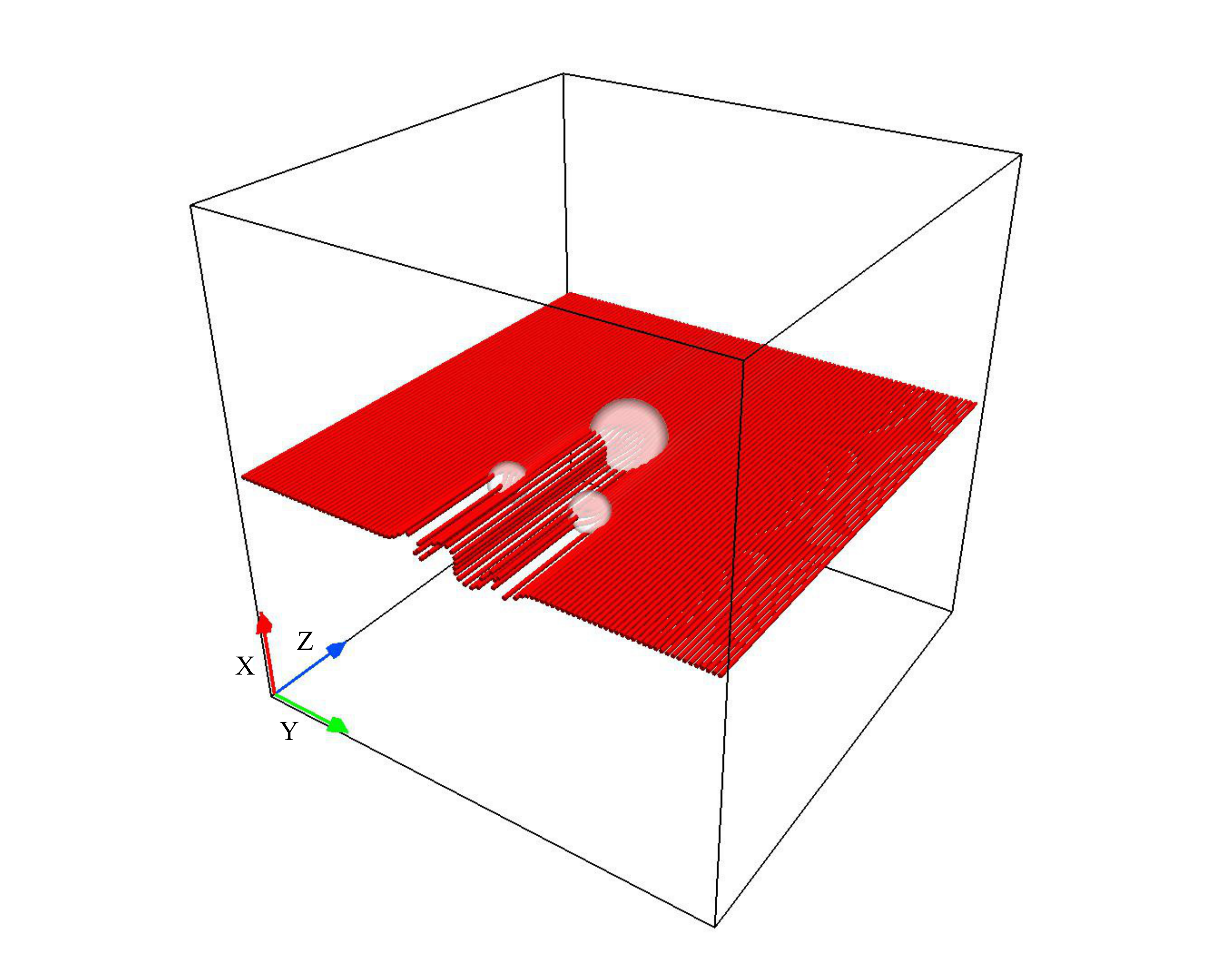}
\caption{Iso-surfaces at $10\%$ of the maximum of $|\mathbf{R}|$, showing the three localized non-ideal regions at $t=1$ in the time dependent model. In red are a selection of field lines plotted from footpoints along $(x,z) = (0,2)$, demonstrating the injection of twist into the field and the overlap of the field line mappings.}
\label{fig:vapor_t1}
\end{figure}

The mapping of field lines on $z=-2$ at $t=1$, color coded according to whether they start above or below $\gamma$ on the other side of the non-ideal region ($z=2$) is shown in Fig. \ref{fig:fluxcount}(a). Figure \ref{fig:fluxcount}(b) shows the regions within which field lines have changed connectivity compared with the mapping at $t=0$. White areas depict where flux has reconnected across $\gamma$ from $x<0$ to $x>0$, and black regions where flux has been reconnected in the other direction. Grey shows regions where field lines have not crossed $\gamma$. Figure \ref{fig:vapor_t1} shows a 3D visualization of the field at $t=1$, were the iso-contours depict the shape and position of each non-ideal region. Applying the flux counting procedure we obtain that
\begin{equation}
\left(\frac{d\Phi}{dt}\right)_{tot} \approx 0.04857, \quad \left(\frac{d\Phi}{dt}\right)_{net} \approx 0.03607
\end{equation}
for a grid of $400^{2}$ starting points. Aside from a small variation due to the discrete nature of the method, these results agree closely with the value obtained by applying Eqns. (\ref{rrtot}) and (\ref{rrnet}).

\begin{figure}
\centering
\includegraphics[width=0.45\textwidth]{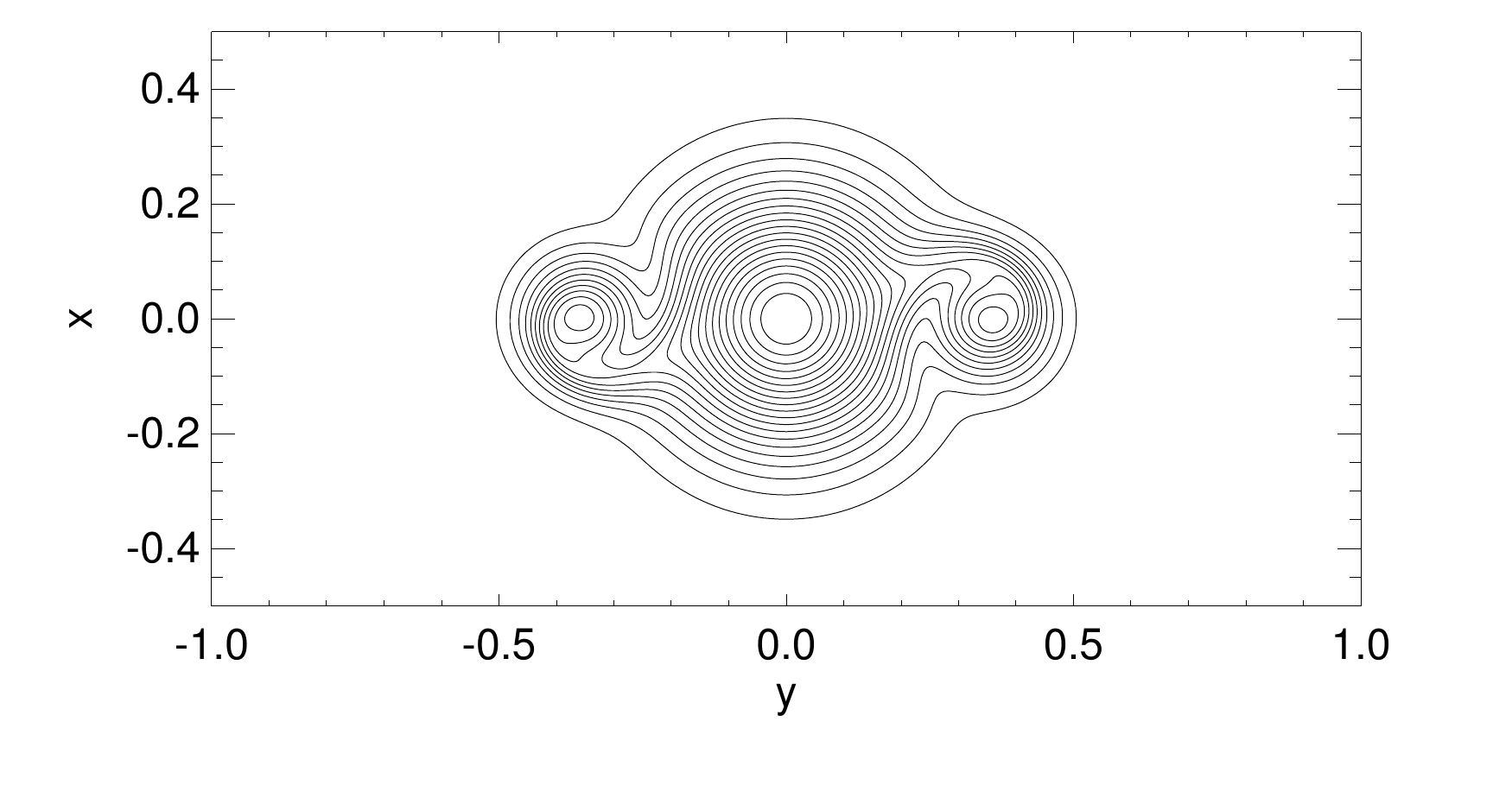}
\caption{The quasi-potential calculated numerically at $t=1$.}
\label{fig:phi_t1}
\end{figure}
 
Lastly, consider now the instantaneous reconnection rate at the later time ($t=1$). At $t=1$ each non-ideal region now adds a non-zero twist to the field line mapping. The overlapping nature of the mappings distorts the shape of $\Xi$ and therefore the positions of the extrema and saddle points, Fig. \ref{fig:phi_t1}. As a result the conceptual flux surface $\gamma$ against which reconnection rate is being measured by Eqn. (\ref{rrtot}) moves to pass through these points at this later time.

\subsection{Steady Sate Reconnection} 
For comparison we now consider the opposite extreme of steady state reconnection for the same initial magnetic field and non-ideal term ($\mathbf{R}$). In steady state the electric field can be expressed in the form of a potential 
\begin{equation}
\mathbf{E} = -\nabla \phi = -\mathbf{v}\times\mathbf{B} + \mathbf{R},
\end{equation}
giving that
\begin{align}
\phi &= -\int{\mathbf{R}\cdot \mathbf{B}/|B| ds }  + \phi_{0}\nonumber \\
&=-\int{E_{\|}ds} + \phi_{0} \nonumber \\
&= \Xi + \phi_{0}
\end{align}
For illustration we set $\phi_{0}=0$ which removes background ideal motions. Thus,
\begin{equation}
\mathbf{E} = -\nabla \Xi. 
\end{equation}
This electric field differs from $\mathbf{R}$, with a non-zero part outside of the non-ideal region which induces a perpendicular plasma flow of the form
\begin{equation}
\mathbf{v}_{\perp} = \frac{\mathbf{E}\times\mathbf{B}}{B^2}.
\end{equation}
The magnetic field in this case remains straight for all time, and the quasi-potential is simply the same as the time dependent case at $t=0$, i.e. 
\begin{equation}
\Xi(x,y) = \sqrt{\pi}\sum_{i=0}^{3} \frac{j_{i}}{l_{z,i}} e^{-(x-x_{0,i})^2/l_{x,i}^2 -(y-y_{0,i})^2/l_{y,i}^2}.
\end{equation}

\begin{figure}
\centering
\includegraphics[width=0.45\textwidth]{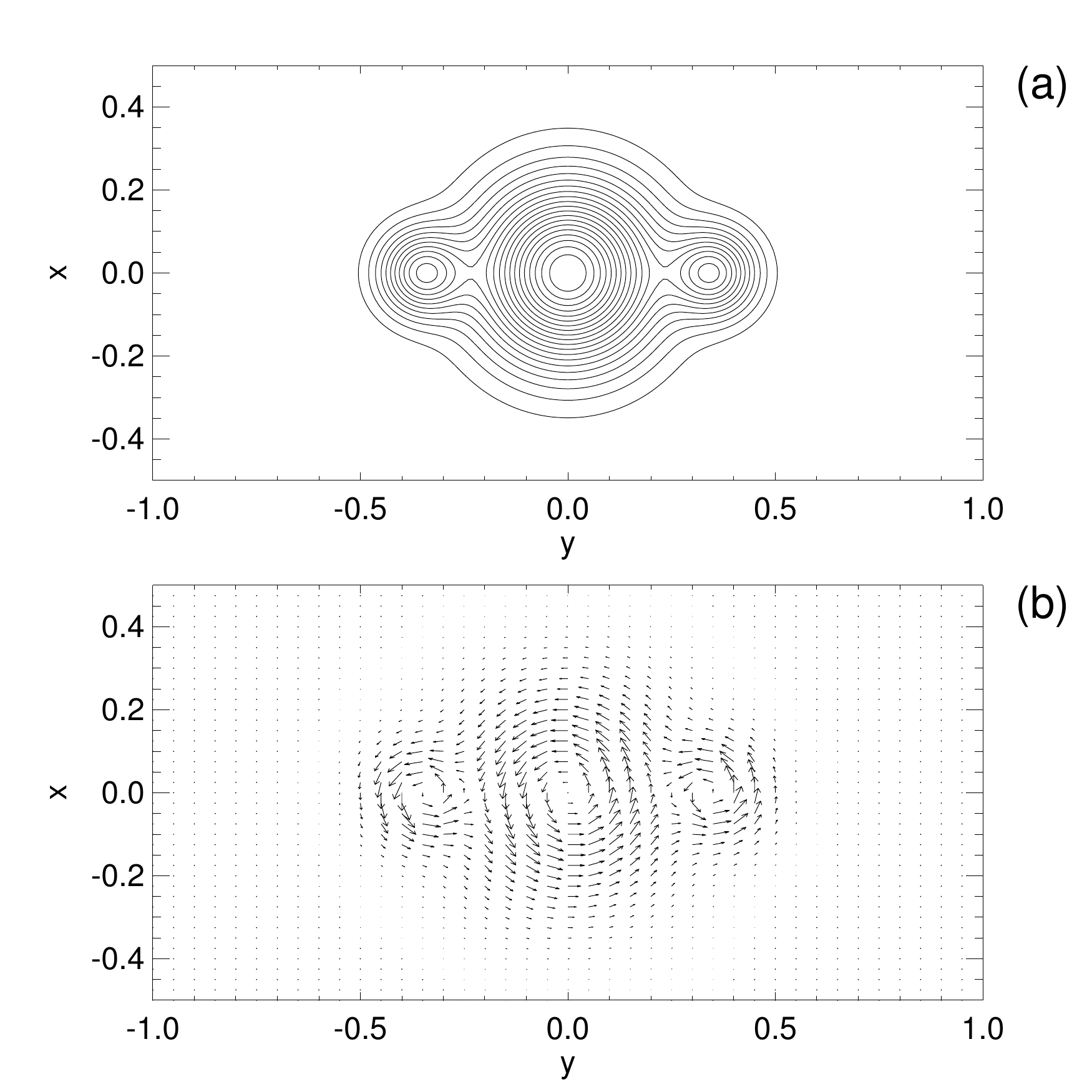}
\caption{(a) $\Xi(x,y)$ in the steady state example. (b) The induced perpendicular plasma flow $\mathbf{v}_{\perp}$ at $z=-2$.}
\label{fig:vel}
\end{figure}

\begin{figure}
\centering
\includegraphics[width=0.45\textwidth]{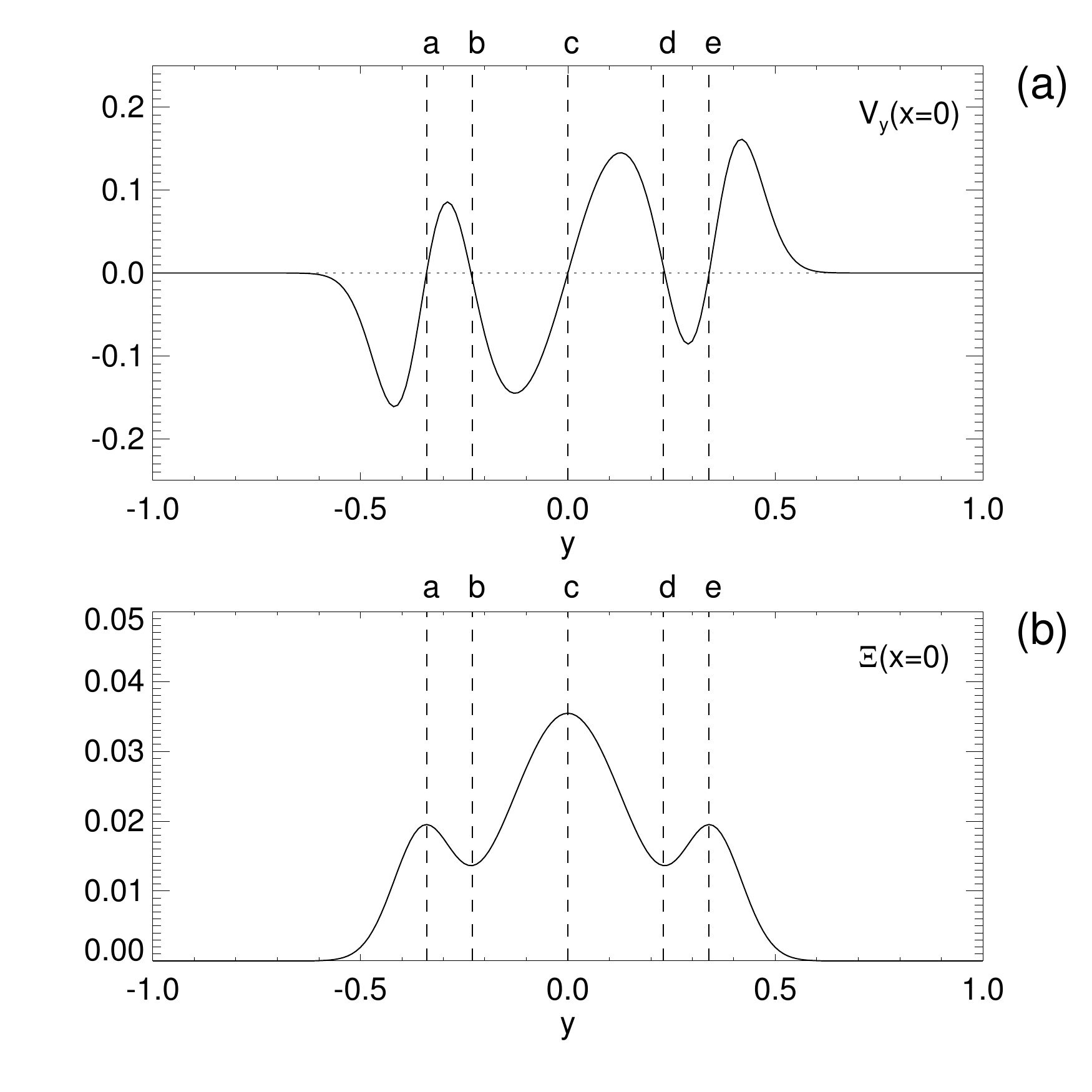}
\caption{(a) velocity perpendicular to the $\gamma$ line ($x=0$) in the steady state example. (b) variation of $\Xi$ along the $\gamma$ line. Note that the zeros in the velocity field correspond to peaks or troughs of $\Xi$.}
\label{fig:vslice}
\end{figure}

Figure \ref{fig:vel} shows the induced plasma flows on one side of the reconnection regions when the electric field is assumed to be zero at $z=2$. The generated flux transporting flows follow the contours of the quasi-potential, producing three overlapping vortices. As the contours of $\Xi$ now form the stream lines of the perpendicular plasma flow, the zeros in the flow pattern are co-located with the peaks and saddle points in $\Xi$, Fig. \ref{fig:vslice}. As the quasi-potential $\Xi$ is the same as the time dependent scenario at $t=0$ the two measures of reconnection rate are then also
\begin{equation}
 \left(\frac{d\Phi}{dt}\right)_{tot}  \approx 0.04711 \quad \& \quad \left(\frac{d\Phi}{dt}\right)_{net} \approx 0.03545,
 \end{equation}
 where $\gamma$ can be chosen to lie along $x=0$. In this extreme these quantities are measures of the total and net rate at which flux is swept past $x=0$ by the induced plasma flow on one side of the collective non-ideal regions, i.e.
\begin{align}
\left(\frac{d\Phi}{dt}\right)_{tot} &= \int_{-\infty}^{\infty} B_{0}  |\mathbf{v}_{\perp}(x=0,z=-2)|_{+/-} \,dy, \label{rrvel1}\\
\left(\frac{d\Phi}{dt}\right)_{net}&= \int_{0}^{\infty} B_{0}  \mathbf{v}_{\perp}(x=0,z=-2) \,dy,
\label{rrvel2}
\end{align}
where $|..|_{+/-}$ denotes integration over either the positive or negative values only.  An approximate expression for this flux transporting flow evaluated on $\gamma$ $(x=0)$ at $z=-2$ is
\begin{equation} 
\mathbf{v}_{\perp, \gamma} = v_{x}(x=0,z=-2) = E_{y}B_{z}/B^2 \approx -\frac{\partial \Xi}{\partial y} B_{0}/B_{0}^2 
\end{equation}
which when substituted into Eqn. (\ref{rrvel1}) and integrated over the regions of negative velocity leads to
\begin{align}
\left(\frac{d\Phi}{dt}\right)_{tot}&=(\Xi_{e} - \Xi_{d}) + (\Xi_{c}-\Xi_{b}) + (\Xi_{a} - 0), \nonumber \\
&= \Xi_{c} + \left|\Xi_{a}-\Xi_{b}\right| + \left|\Xi_{e}-\Xi_{d}\right|. 
\label{phitot}
\end{align} 
Note that integrating over the positive value gives the same result. Substituting the above expression for $\mathbf{v}_{\perp, \gamma}$ into Eqn. (\ref{rrvel2}) then also gives that
\begin{equation}
\left(\frac{d\Phi}{dt}\right)_{net} = \Xi_{c}.
\label{phinet}
\end{equation} 
Eqns. (\ref{phitot}) and (\ref{phinet}) are simply Eqns. (\ref{rrtot}) and (\ref{rrnet}) applied to this particular $\Xi$ profile. 

Thus, we have verified the two rates of reconnection for our idealized fragmented reconnection region in each of the two extreme cases of steady state and purely time dependent reconnection and by extension the continuum of cases in-between.

\section{Discussion and Conclusions}
The aim of this paper was to extend the theory of General Magnetic Reconnection to situations with fragmented current layers within a localized volume. We considered the manner in which new connections may be formed, derived expressions for the rate at which this occurs and verified these expressions with two simple examples.

In terms of facilitating the formation of new connections we showed that in the extreme of steady state reconnection a large scale rotational non-ideal flow with internal vortices is produced, whilst purely time dependent reconnection leads to spontaneously braided magnetic fields. However, it should be emphasized that the reverse is also true.  That is, the existence of non-ideal regions is guaranteed by the right evolution of the magnetic field (given the necessary non-ideal plasma conditions). In particular, if a magnetic field is initially braided with the field lines entering and leaving the volume held fixed, then multiple current layers must form to remove this braiding. This second scenario is readily observed by numerical experiments examining the non-ideal relaxation of braided magnetic fields (e.g. \citet{Pontin2011b,Rappazzo2013}).

By considering the closed paths along which these new connections formed we also showed that when current layers are fragmented {\it{two rates of reconnection}} can be defined which describe the process. $(d\Phi/dt)_{tot}$ which measures the true rate at which new connections are formed collectively by the multiple non-ideal regions and a second, $(d\Phi/dt)_{net}$ measuring the net rate at which changes in the global field occurs. When applied to a single reconnection region both rates are equal. 

We chose to define $(d\Phi/dt)_{tot}$ such that it measures the total rate at which flux is locally and globally cycled when viewed in flux coordinate space. This requires evaluating the quasi-potential at the saddle points of $\Xi$ as well as the extrema. We chose this rather than a simple sum over each extrema as summing over only the extrema overestimates the rate flux is cycled (although if each non-ideal region has little overlap this may give a close approximation, e.g. \citet{Pontin2011b}). This occurs as each extrema taken on its own measures the net rate of transfer of flux between itself and the background ideal field. Therefore, summing over all extrema double counts the flux being cycled around outer loops, such as those depicted in orange and yellow in Fig. \ref{fig:arc}. By involving the quasi-potential measured at the saddle points, this double counting is avoided. 

It is also worth emphasizing that our total reconnection rate $(d\Phi/dt)_{tot}$ does not measure the sum of the reconnection rates of each individual reconnection region within the volume. The only way that this could be quantified would be to consider the local quasi-potential drop across each non-ideal region in turn. However, each region would have to be surrounded by ideal magnetic field for this to be meaningful. In fragmented current layers this is rarely the case as different current sheets partially overlap when merging or breaking apart. Considering the collective behavior as we have done here is the only way to properly quantify such a system.

Given that we have introduced two different rates to describe this collective behavior, which should be used to characterise a given reconnection process? It depends upon what is most of interest for the problem at hand. For instance, if one is considering the scaling of energy release compared with reconnection rate then the total rate is the better choice. It would also be the more relevant choice in situations  where the rate at which flux is swept up by a fragmented reconnection region is of interest, as is thought to be related to photospheric brightening in solar flares (e.g. \citet{Qiu2010,Hesse2005}). However, the net rate may be more useful when the multiple reconnection regions are fluctuating and transient (as occurs during an increasing turbulent evolution of the magnetic field) and there are some simple large scale symmetries against which flux transfer is wished to be know (e.g. \citet{Wendel2013b,Kowal2009}).

Ultimately the non-ideal physics associated with  the plasma, any gradients in the mapping of the magnetic field and the way in which excess magnetic energy is built up will dictate where non-ideal regions form and if they subsequently fragment. The present analysis serves as a way of interpreting how the subsequent reconnection proceeds and how best to quantify it.

\begin{acknowledgements}
This research was supported by NASA's Magnetospheric Multiscale mission. PW acknowledges support from an appointment to the NASA Postdoctoral Program at Goddard Space Flight Center, administered by Oak Ridge Associated Universities through a contract with NASA. Figs. 9 and 12 were made using the Vapor visualization package (www.vapor.ucar.edu).
\end{acknowledgements}


%

\end{document}